\PassOptionsToPackage{dvipsnames}{xcolor}	
\documentclass[12pt,letter]{article}

\usepackage[T1]{fontenc}
\usepackage{lmodern}
\usepackage{setspace}
\usepackage{etoolbox}
\makeatletter
\patchcmd{\appendix}{\@Alph}{\@Roman}{}{}
\makeatother

\usepackage{amsmath}
\usepackage{amssymb}
\usepackage{amsthm}
\usepackage{mathtools}
\usepackage{mathrsfs}
\usepackage{breqn}
\usepackage{bigints}

\usepackage[top=1.25 in, bottom=1.25 in, left= 1 in, right=1 in]{geometry}
\usepackage{enumerate}
\usepackage{enumitem}
\setlist[enumerate,1]{label=(\arabic*)}
\setlist[itemize,1]{label=--}    
\usepackage{xcolor}
\usepackage{hyperref}
\usepackage{float}
\usepackage{indentfirst}
\usepackage{caption}
\usepackage{tikz}
\usetikzlibrary{decorations.pathreplacing}
\usepackage{mathtools}
\usetikzlibrary{positioning}
\usepackage{graphicx}
\usetikzlibrary{positioning,chains,fit,shapes,calc}
\usepackage{tkz-berge}
\usetikzlibrary{fit,shapes}

\newcommand{\bb}{\mathbb}

\newcommand{\A}{\bigcap}
\newcommand{\und}{\underline}
\newcommand{\lims}{\lim\limits}

\newcommand{\supp}{\text{supp}}

\renewcommand{\epsilon}{\varepsilon}

\newtheorem{theorem}{Theorem}
\newtheorem{theorem0}{Theorem}
\newtheorem{lemma}{Lemma}
\newtheorem{proposition}{Proposition}
\newtheorem{corollary}{Corollary}
\newtheorem{assumption}{Assumption}

\newtheorem*{statement*}{Result}

\newtheorem{remark}{Remark}

\theoremstyle{definition}
\newtheorem{definition}{Definition}

\DeclareMathOperator*{\argmax}{\arg\!\max}

\DeclareTextFontCommand{\emph}{\slshape}

\usepackage{natbib}
\nocite{*}

\title{Marginal Reputation}
\author{Daniel Luo and Alexander Wolitzky \\
MIT\thanks{For helpful comments, we thank seminar participants at ESWC, Northwestern, UT Austin, and VSET; and Sandeep Baliga, Ian Ball, V.\ Bhaskar, Drew Fudenberg, Eric Gao,  Navin Kartik, Anton Kolotilin, Stephen Morris, David Pearce, Harry Pei, and Tom Wiseman.}}
\date{\today}

\begin{document}

\onehalfspacing
\maketitle

\begin{abstract}
    We study reputation formation where a long-run player repeatedly observes private signals and takes actions. Short-run players observe the long-run player's past actions but not her past signals. The long-run player can thus develop a reputation for playing a distribution over actions, but not necessarily for playing a particular mapping from signals to actions. Nonetheless, we show that the long-run player can secure her Stackelberg payoff if distinct commitment types are statistically distinguishable and the Stackelberg strategy is \emph{confound-defeating}. This property holds if and only if the Stackelberg strategy is the unique solution to an optimal transport problem. If the long-run player's payoff is supermodular in one-dimensional signals and actions, she secures the Stackelberg payoff if and only if the Stackelberg strategy is monotone.  Applications include deterrence, delegation, signaling, and persuasion.  Our results extend to the case where distinct commitment types may be indistinguishable but the Stackelberg type is \emph{salient} under the prior. 
\end{abstract}

\textbf{Keywords.} Reputation, repeated games, confound-defeating, optimal transport, cyclical monotonicity, deterrence, delegation, signaling, Bayesian persuasion. \bigskip

\textbf{JEL Codes.} C73, D83

\thispagestyle{empty}

\newpage
\pagenumbering{arabic} 

\section{Introduction}

This paper considers reputation formation in settings where one desires a reputation not only for taking certain actions, but for acting in the right circumstances. Our main applications are to deterrence, delegation, and communication games, where the importance of establishing a reputation for \emph{conditional} action has long been accepted in the informal literature. For example, \cite{schelling1966arms} writes,
\begin{quote}
    ``any coercive threat requires corresponding \emph{assurances}; the object of a threat is to give somebody a choice. To say, ``One more step and I shoot,'' can be a deterrent threat only if accompanied by the implicit assurance, ``And if you stop I won't.'' Giving notice of \emph{unconditional} intent to shoot gives him no choice.''
\end{quote}
Similarly, when an informed sender asks a receiver to take an action that the sender prefers, the request is persuasive only if the receiver believes that the sender tends to make it only when compliance is in the receiver's interest.

To study reputation formation in these settings, we consider a model where a long-run player facing a sequence of short-run opponents repeatedly observes private signals and takes actions. For example, in the deterrence context, the signal is whether the long-run player detects an attack by a short-run player who moves first, and the action is whether she fights back. In the communication context, the long-run player is the first-mover, the signal is a payoff-relevant state variable, and the action is a signal or message to a short-run player who moves second. The long-run player is either rational or is one of a number of possible \emph{commitment types} that play a fixed mapping from signals to actions in each period. The set of possible commitment types includes the \emph{Stackelberg type} that plays the long-run player's most-preferred commitment strategy.

In this setup, if short-run players observe the history of the long-run player's past actions \emph{and signals}, standard results imply a patient long-run player is assured at least her Stackelberg (best commitment) payoff in every Nash equilibrium (\citeauthor{FudenbergLevine89} \citeyear{FudenbergLevine89,FudenbergLevine92}). We instead consider the case where the long-run player's actions are observed, but her signals are not. In the deterrence context, this says that potential attackers know when the long-run player has fought in the past, but not whether this fighting came in response to detected attacks. In the communication context, it says that receivers observe the history of messages sent by the long-run player, but not the history of states.

Existing results say little about the outcomes of these games. The key issue is that the long run player's strategy---how she maps signals to actions---is not identified by the observed marginal distribution over her actions. This implies that existing payoff bounds for reputation games with imperfect monitoring \citep{FudenbergLevine92,Gossner11} are extremely loose and often trivial in our setting. For example, suppose that in the deterrence context the long-run player follows her (pure) Stackelberg strategy of fighting if and only if she detects an attack. This strategy results in the long-run player fighting a certain fraction of the time, say 50\%. However, after seeing her fight half the time, potential attackers need not come to believe that she is playing the Stackelberg strategy---they might instead believe that she is playing a different strategy with the same marginal over actions, such as fighting half the time independent of her signal. Which inference potential attackers draw is critical for the long-run player, as they will be deterred if they believe she is playing the Stackelberg strategy of fighting when she detects an attack, but not if they believe she is randomly fighting half the time. As existing results do not restrict the short-run players' inferences in this situation, they make no non-trivial predictions about the long-run player's payoff. Formally, in this example attacking and not attacking are both ``0-confirmed best responses'' to the Stackelberg strategy---meaning that they are both best responses to some long-run player strategy that induces the same signal distribution as the Stackelberg strategy---which implies that the payoff lower bound of \cite{FudenbergLevine92} is vacuous.

Our main result provides conditions for a patient long-run player to secure her Stackelberg payoff when only the marginal over actions is identified. The key sufficient condition is that the Stackelberg strategy is \emph{confound-defeating}: against any 0-confirmed best response, the Stackelberg strategy is uniquely optimal among strategies that induce the same marginal over actions. Intuitively, if the Stackelberg strategy is confound-defeating then the rational long-run player never plays a different strategy that induces the same marginal in any Nash equilibrium. Therefore, establishing a reputation for playing the ``Stackelberg marginal'' suffices to establish a reputation for playing the Stackelberg strategy. (Our theorem also requires that the Stackelberg strategy is not confounded by a different commitment type of the long-run player---we discuss this second condition below.)

A strategy is confound-defeating if and only if the induced joint distribution over the long-run player's signal and action is uniquely optimal among all distributions with the same marginals: that is, if and only if it is the unique solution to the optimal transport problem of maximizing the long-run player's payoff subject to given marginals over her signal and action. Adapting standard optimal transport results, we show that this holds if and only if the support of the induced joint distribution satisfies a strict version of cyclical monotonicity \citep{rochet1987necessary}. We also use cyclical monotonicity to provide a converse to our main result: if the long-run player is rational with high probability, her payoff in any equilibrium cannot exceed that from some cyclically monotone strategy.

The cyclical monotonicity characterization makes confound-defeatingness easy to check in many games. In particular, if the long-run player's payoff is strictly supermodular in one-dimensional signals and actions, a strategy is confound-defeating if and only if every selection from its support is monotone. Applied to the deterrence game, this says that the long-run player can secure her Stackelberg payoff if fighting is relatively more appealing when an attack is detected. Conversely, if the long-run player is rational with high probability, her payoff cannot exceed that from a monotone strategy in any Nash equilibrium. For example, in the deterrence game, the long-run player obtains close to her minmax payoff in every Nash equilibrium if fighting is more appealing when an attack is not detected. Combining these results gives a unique payoff prediction in many games when the long-run player is patient and is rational with high probability.

Our results have strong implications for repeated communication games. An immediate implication is that, in repeated signaling games where the sender's payoff is additively separable in the receiver's action and strictly supermodular in her own (one-dimensional) type and action, a patient sender can secure her best commitment payoff from any monotone signaling strategy.\footnote{The results of \citet{FudenbergLevine92} imply the sender can secure her Stackelberg payoff in repeated signaling games where actions \emph{and states} are observed at the end of each period. Our results imply the same conclusion holds when only actions are observed, if the Stackelberg strategy is monotone and the sender's payoff is additively separable in the receiver's action and supermodular in her own type and action.} This follows because these conditions imply that the sender's payoff is strictly supermodular in her own type and action for any receiver strategy, so our results for one-dimensional supermodular games apply.

We then consider repeated cheap talk games with state-independent sender preferences. Our results do not apply to cheap talk games, because when the sender's ``action'' is a payoff-irrelevant message, her payoff cannot be strictly supermodular in this action and her type. However, our results imply that perturbing the sender's payoff by adding a small, strictly submodular ``lying cost'' can provide a reputational foundation for any communication mechanism that is monotone with respect to some order on states and receiver actions. While not fully general, this class includes all partitions (deterministic communication mechanisms) and all linear partitions with randomization at the boundaries. We can thus provide a reputational foundation for a large class of communication mechanisms, even when the history of realized states is unobserved.

Our main result assumes that distinct commitment types in the support of the short-run players' prior are statistically distinguishable. Without this assumption, non-Stackelberg play by commitment types with the same marginal as the Stackelberg strategy can hinder reputation formation. Nonetheless, we show that our result goes through when the Stackelberg type is \emph{salient} under the prior. Roughly, this condition says that the Stackelberg type has sufficiently high prior weight relative to other commitment types that induce the same marginal but different best responses. In the deterrence context, this says that short-run players believe that, conditional on the long-run player being irrational, she is more much likely to play the strategy ``fight if and only if an attack is detected'' than the strategy ``fight half the time independent of the signal.''

\paragraph{Related Literature.} 
We contribute to the literature on reputation formation with imperfect monitoring, introduced by \cite{FudenbergLevine92}. They show that a patient long-run player can secure her commitment payoff against the least favorable of her opponent's $0$-confirmed best responses. 
In our partially identified setting, the set of $0$-confirmed best responses is typically large, so this payoff lower bound is weak and often vacuous. \cite{Gossner11} gives a different proof---which we build on---of a similar lower bound, which is also too weak in our setting for the same reason. \citeauthor{FudenbergLevine92} and \citeauthor{Gossner11} also give upper bounds for a patient long-run player's payoff, which \cite{ElyValimaki03} show is much too loose in a class of repeated delegation games. In contrast, we show that their lower bounds are much too loose in a class of games where the long-run player observes private signals and the Stackelberg strategy is confound-defeating.\footnote{\cite{ely2008reputation} add ``good commitment types'' to \cite{ElyValimaki03} and show that this does not restore the long-run player's commitment payoff. The reason for the difference from our results is that we assume that the long-run player's action is always identified, whereas in \cite{ely2008reputation} it is not identified when the short-run player exits. See Section \ref{s:delegation}.} 

\cite{pei2020interdependent} studies a reputation model with interdependent values, where a possibly committed long-run player privately observes a perfectly persistent, payoff-relevant state. Our model instead covers (as a special case) the case where the state is i.i.d.: see Section \ref{s:trust}. In both papers, supermodularity-type conditions are important for securing the Stackelberg payoff, but the precise conditions and arguments are very different.\footnote{For example, \citeauthor{pei2020interdependent}'s Stackelberg payoff theorem requires binary actions for the short-run player and a condition on the prior, while we require no such conditions.} Other papers in the reputation literature where supermodularity conditions play key roles in deriving payoff bounds include \cite{liu2011information}, \cite{liu2014limited}, and \cite{pei2024shortmemories}.

We also relate to a diverse literature on games and mechanisms where strategies are partially identified, so certain deviations are undetectable. The connection between incentive compatibility and optimal transport in such settings dates to \cite{rochet1987necessary}; \cite{rahman2024detecting} gives an alternative interpretation and proof. Applications include quota mechanisms \citep{jackson2007overcoming,matsushima2010role,escobar2013efficiency,frankel2014aligned,ball2024quota}, multidimensional or repeated cheap talk \citep{chakraborty2007comparative,renault2013dynamic,margaria2018dynamic,meng2021value}, and repeated random matching games \citep{takahashi2010community,heller2018observations,clark2021record}. A particularly related paper is \citet{lin2024credible}, who study optimal information disclosure via cheap talk when the marginal distribution over messages is observed. In their setting, a joint distribution over states and receiver actions is implementable if it maximizes the sender's payoff over all joint distributions with the same marginals. In contrast, our confound-defeating property requires the joint distribution over states and \emph{sender} actions to uniquely maximize the sender's payoff over all joint distributions with the same marginals, for any receiver best response. The two conditions are thus related, but involve different objects (distributions over states and receiver actions vs. payoff-relevant sender actions) and come from different strategic considerations (static cheap talk subject to a ``credibility'' constraint vs. long-run reputation formation).\footnote{\cite{renault2013dynamic} characterize the set of equilibrium payoffs in repeated cheap talk games with patient players via an optimality condition over joint distributions of states and receiver actions similar to that in \cite{lin2024credible}.} 

We are also not the first to discuss reputation foundations for the Bayesian persuasion commitment assumption. This assumption has been controversial ever since its introduction by \cite{KamenicaGentzkow11} and \cite{rayo2010persuasion}, who suggested reputation as a possible foundation. \cite{mathevet2024reputation} observe that reputation effects yield the commitment payoff when a long-run sender faces a sequence of short-run receivers who observe the history of messages \emph{and states}, and they study whether the sender's \emph{behavior} likewise coincides with the commitment solution. We instead ask when reputation effects yield the commitment payoff when receivers observe past messages but not states. Farther afield, \cite{pei2023privateLying} and \cite{best2024persuasion} show that the commitment payoff arises as one of many equilibrium payoffs in repeated communication game with lying costs and coarse information on histories, respectively; \cite{kuvalekar2022goodwill} study repeated sender-receiver games with two long-run players and unobserved past states, without commitment types; and \cite{fudenberg2022reputation} provide a foundation for the commitment payoff in a model where the long-run player sends messages before taking actions and can develop a reputation for ``honesty'' about the action they are about to take.

\paragraph{Organization.} Section \ref{s:deterrence} analyzes two examples, including the deterrence game discussed above. Section \ref{s:model} develops the general model. Section \ref{s:general} presents the confound-defeating property and our main result: the long-run player can secure her Stackelberg payoff if the Stackelberg strategy is confound-defeating and not behaviorally confounded. Section \ref{s:cyclical} characterizes confound-defeatingness via cyclical monotonicity and gives an upper bound for the long-run player's payoff. Section \ref{s:monotone} applies our results to one-dimensional supermodular games. Section \ref{s:commmunication} considers communication games. Section \ref{s:salience} extends our results to allow indistinguishable commitment types by introducing our salience notion. Section \ref{s:discussion} summarizes the paper and discusses some possible extensions.

\section{Motivating Examples} \label{s:deterrence}

\subsection{Deterrence with Private Signals}
We begin with the example of a simple deterrence game.\footnote{The stage game in this subsection is an example of an \emph{inspection game} (\cite{avenhaus2002inspection}). \cite{acemoglu2023mistrust} survey deterrence and related games in economics and political science, emphasizing the role of private signals.} A long-run player with discount factor $\delta$ faces a sequence of short-run opponents. Each period, the short-run player first chooses whether to \emph{Cooperate} ($C$) or \emph{Defect} ($D$). The long-run player then observes a private signal, $c$ or $d$, drawn with conditional probability $\Pr(c|C)=\Pr(d|D)=p \in (1/2,1)$, before choosing whether to \emph{Accommodate} ($A$) or \emph{Fight} ($F$). The short-run player observes the history of the long-run player's past actions, but not her signals. The short-run player's payoff is specified as a function of both players' actions while the long-run player's payoff is specified as a function of her action and private signal as follows.\footnote{An interpretation of the dependence of the long-run player's payoff on her signal rather than the short-run player's action is that each period's short-run player is ``crazy'' with probability $2(1-p)$, independently across periods, in which case she mixed 50-50 between her actions, and this is the only source of noise. In this case, the long-run player's ``signal'' is just the short-run player's action, accounting for the crazy types.}

\begin{figure}[h!]
    \centering
    \begin{minipage}{0.5\linewidth}
        \centering
        \[
        \begin{array}{c c c c}
            & C & D & \\ 
                        A & 1 & 1+g & \\
                        F & -l & 0 & \\
                    \end{array}
        \]
        \vspace{-1.5em}
        \captionof{table}{Short-Run Player Payoff}
    \end{minipage}%
    \hspace{-5em}
    \begin{minipage}{0.5\linewidth}
        \centering
        \[
        \begin{array}{c c c c}
            & c & d & \\ 
                       A & 1 & y & \\
                      F & x & 0 & \\
                   \end{array}
        \]
        \vspace{-1.5em}
        \captionof{table}{Long-Run Player Payoff}
    \end{minipage}
\end{figure}

Assume that $g,l>0$ and $x,y \in (0,1)$, so that $D$ is dominant for the short-run player and $A$ is dominant for the long-run player. Assume also that $p$ is sufficiently large that the short-run player strictly prefers to take $C$ if the long-run player plays the ``deterrence'' strategy, \emph{(A after c; F after d)}, which we denote as $(A,F)$; and that the long-run player would rather play $(A,F)$ against $C$ than play $(A,A)$ against $D$, so that $(A,F)$ is the long-run player's pure Stackelberg strategy.\footnote{These two conditions holds iff $p>\max \{\frac{1+g+l}{2+g+l},\frac{1}{2-y}$\}.} Finally, assume that the long-run player is committed to $(A,F)$ with probability $\mu_0$ and is rational otherwise.\footnote{The role of this assumption in the context of the current example in discussed in Sections \ref{s:confounding} and \ref{s:salience}.}

What can be said about the equilibria of this game when the discount factor $\delta$ is close to $1$? A first observation is that, in \citeauthor{FudenbergLevine92}'s (\citeyear{FudenbergLevine92}) terminology, both $C$ and $D$ are ``$0$-confirmed best responses'' to the Stackelberg strategy $(A,F)$. For example, if the short-run player takes $D$, the long-run player ends up fighting with probability $p$ when she plays $(A,F)$, but also when she fights with probability $p$ after each signal. Since $D$ is a best response to the latter strategy, it is also a $0$-confirmed best response to $(A,F)$.

The results of \cite{FudenbergLevine92} (Theorem 3.1) and \cite{Gossner11} (Corollary 1) state that, as $\delta \to 1$, the long-run player's payoff in any Nash equilibrium is at least her payoff from playing the Stackelberg strategy against the least favorable $0$-confirmed best response. (See Theorem \ref{theorem0: fl92} in Section \ref{s:repeated}.) Since $D$ is a $0$-confirmed best response to $(A,F)$, these results say only that the long-run player's payoff is above $1-p$. When noise is small, so that $p$ is close to $1$, this just says that the long-run player's payoff is feasible.\footnote{\citeauthor{FudenbergLevine92} and \citeauthor{Gossner11} also give payoff upper bounds in terms of the mixed Stackelberg action. Our general results similarly allow mixed commitment types.}

In contrast, we have the following result. Note that the long-run player's pure Stackelberg payoff is $p$, while her minmax payoff is $1-p+py$.

\begin{proposition} \label{p:deterrence}
    Let $\underline{U}_1(\delta)$ and $\bar{U}_1(\delta)$ be the infimum and supremum of the long-run player's payoff in any Nash equilibrium. The following hold:
    \begin{enumerate}
        \item If $x+y<1$ then $\liminf_{\delta \to 1} \underline{U}_1(\delta) \geq p$ for all $\mu_0>0$.
        \item If $x+y>1$ then $\lim_{\mu_0 \to 0} \overline{U}_1(\delta) = 1-p+py$  for all $\delta <1$.
    \end{enumerate}
\end{proposition}

That is, if $x+y<1$, a patient long-run player is assured at least her Stackelberg payoff in any Nash equilibrium; while if $x+y>1$, the long-run player obtains close to her minmax payoff whenever the prior commitment probability is small.

To see why it matters whether $x+y$ is below or above $1$, note that the long-run player's payoff is strictly supermodular in her signal and action in the first case (with the order $A \succ F$, $c \succ d$) and is strictly submodular in the second. In the strictly supermodular case, $F$ is relatively more appealing when $d$ is observed. This implies that the Stackelberg strategy $(A,F)$ strictly outperforms any other strategy that induces the same marginal over actions, and hence is ``confound-defeating.'' As we show in Theorem \ref{main payoff commitment theorem}, this implies that establishing a reputation for playing the Stackelberg marginal over actions suffices to establish a reputation for playing the Stackelberg strategy, and hence secures the Stackelberg payoff.\footnote{This also relies on the assuming that the Stackelberg strategy is not confounded by another commitment type. For example, if there is a second commitment type that fights with probability $p$ after each signal, the conclusion of Proposition \ref{p:deterrence}.1 does not hold.}

Conversely, in the  strictly submodular case, $F$ is relatively more appealing when $c$ is observed. This implies that the rational long-run player always plays $F$ with (weakly) higher probability when $c$ is observed, and plays $A$ with higher probability when $d$ is observed. Such a strategy \emph{encourages} the short-run player to take $D$ rather than deterring him, so the short-run player must take $D$ whenever he believes that the long-run player is rational with high probability. Finally, since beliefs are martingale, it follows that the short-run player usually takes $D$ when the ex ante commitment probability is small.\footnote{The submodular case of Proposition \ref{p:deterrence} follows from Corollary \ref{c:monotone} in Section \ref{s:monotone}.}

\subsection{A Repeated Trust Game} \label{s:trust}

For another example, suppose that the stage game is the following ``trust game'' (or ``product choice game''), adapted from \citet{pei2020interdependent}. There is a \emph{state} $\theta \in \{\text{G(ood), B(ad)}\}$, drawn i.i.d.\ across periods with equal probability on each state. In each period, the long-run player observes $\theta$ before taking an action $a_1 \in \{\text{H(igh Effort), L(ow Effort)}\}$. Simultaneously (and having observed the history of past actions, but not states), the short-run player takes an action $a_2 \in \{\text{T(rust), N(ot Trust)}\}$. Payoffs in each state are given by the following matrices, with the long-run player's payoff listed first in each entry.

\begin{figure}[h!]
    \centering
    \begin{minipage}{0.5\linewidth}
        \centering
        \[
        \begin{array}{c c c c}
            & T & N & \\ 
                        H & 1,2 & -1,0 & \\
                        L & 2,-1 & 0,0 & \\
                    \end{array}
        \]
        \vspace{-1.5em}
        \captionof{table}{Payoffs in State $\theta=G$}
    \end{minipage}%
    \hspace{-5em}
    \begin{minipage}{0.5\linewidth}
        \centering
        \[
        \begin{array}{c c c c}
            & T & N & \\ 
                       H & 1-w,-1 & -1-z,0 & \\
                      L & 2,-1 & 0,0 & \\
                   \end{array}
        \]
        \vspace{-1.5em}
        \captionof{table}{Payoffs in State $\theta=B$}
    \end{minipage}
\end{figure}

Assume that $w,z>-1$, so that $L$ is dominant for the long-run player, and the unique stage-game Nash equilibrium outcome is $(L,N)$ in both states. Note that $T$ is optimal for the short-run player only if he believes that, with high enough probability, \emph{both} $\theta=G$ and $a_1=H$. For example, suppose that player 1 is the chef of a seafood restaurant, $\theta$ is the quality of the day's catch, and player 2 is a customer who wants to eat only fish that is both of high quality and carefully cooked. Note that the long-run player's pure Stackelberg strategy is $(H,L)$ (as this strategy lets the long-run player enjoy taking $L$ in state $B$ while inducing the short-run player to take $T$), and assume that she is committed to this strategy with probability $\mu_0$. Note also that the long-run player's pure Stackelberg payoff is $3/2$, while her minmax payoff is $0$.

\begin{proposition} \label{p:trust}
The following hold:
    \begin{enumerate}
        \item If $\min\{w,z\}>0$ then $\liminf_{\delta \to 1} \underline{U}_1(\delta) \geq 3/2$ for all $\mu_0>0$.
        \item If $\max\{w,z\}<0$ then $\lim_{\mu_0 \to 0} \overline{U}_1(\delta) = 0$ for all $\delta <1$.
    \end{enumerate}
\end{proposition}

While the timing of the deterrence and trust games are different (e.g., the short-run player moves first in the former and simultaneously with the long-run player in the latter), the logic of Proposition \ref{p:trust} is similar to that of Proposition \ref{p:deterrence}. If $\min\{w,z\}>0$ then the long-run player's payoff is strictly supermodular in $(\theta,a_1)$ for any $a_2$, which we show lets her secure her Stackelberg payoff when she is patient. If instead $\max\{w,z\}<0$ then the long-run player's payoff is strictly submodular in $(\theta,a_1)$ for any $a_2$, which we show limits her to her minmax payoff when the prior commitment probability is small.\footnote{If one of $\{w,z\}$ is positive and the other negative, the long-run player's payoff is supermodular in $(\theta,a_1)$ for one $a_2\in \{T,N\}$ and submodular for the other. Our results do not cover this case. We also note that Proposition \ref{p:trust} is roughly consistent with \citeauthor{pei2020interdependent}'s (\citeyear{pei2020interdependent}) results for the case where $\theta$ is perfectly persistent: \citeauthor{pei2020interdependent} shows that the long-run player can fail to secure her Stackelberg payoff in the submodular product choice game, but does secure it in the supermodular case (under an additional condition on the prior).}

\section{Model} \label{s:model}

We study repeated games where a possibly-committed long-run player faces a sequence of short-run opponents. To cover both applications where a short-run player moves first (like deterrence games) and those where a short-run player moves simultaneously with or after the long-run player (like trust or communication games), we consider repeated three-player games, where one short-run player (player 0) moves first, and then the long-run player (player 1) and another short-run player (player 2) move simultaneously. In deterrence games, the second short-run player is absent. In trust, delegation, and communication games, the first short-run player is Nature. In addition, in communication games, the second short-run player's action is a mapping from the long-run player's action to a set of possible responses.

We first describe the stage game, followed by the repeated game.

\subsection{The Stage Game}
There are three players, $i \in \{0, 1, 2\}$. Player 1 is the long-run player; players 0 and 2 are short-run players. Each player $i$ has a finite action set $A_i$ with generic element $a_i$. There are also two finite signal sets, $Y_0$ and $Y_1$, with generic elements $y_0$ and $y_1$.

The stage game timing is as follows.

\begin{enumerate}
    \item Player $0$ takes an action $a_0$. This generates a signal $y_0$, drawn from a distribution $\rho_0(\cdot | a_0)$, which is observed by player $1$ only.
    \item Players $1$ and $2$ simultaneously take actions $a_1$ and $a_2$. This generates a signal $y_1$, drawn from a distribution $\rho_1(\cdot | a_1, a_2)$, which is publicly observed by all players.
\end{enumerate}

Thus, stage game strategies for players $0$ and $2$ are simply mixed actions $\alpha_0 \in \Delta(A_0)$ and $\alpha_2 \in \Delta(A_2)$, respectively, while a stage game strategy for player $1$ is a function $s_1: Y_0 \to \Delta(A_1)$. Note a strategy profile $(\alpha_0, s_1, \alpha_2)$ induces a joint distribution $\gamma(\alpha_0,s_1) \in \Delta(Y_0 \times A_1)$ (independent of $\alpha_2$) over player $1$'s private signal $y_0$ and action $a_1$ according to
\[ \gamma(\alpha_0, s_1)[y_0, a_1] = \sum_{a_0 \in A_0} \alpha_0(a_0)\rho_0(y_0 | a_0)s_1(y_0)[a_1], \]
and induces a joint distribution $p(\alpha_0, s_1, \alpha_2) \in \Delta(Y_1)$ over the public signal $y_1$ according to
\[ p(\alpha_0, s_1, \alpha_2)[y_1] = \sum_{a_0 \in A_0} \sum_{y_0 \in Y_0} \sum_{a_1 \in A_1} \sum_{a_2 \in A_2} \alpha_0(a_0) \rho_0(y_0 | a_0) s_1(y_0)[a_1] \alpha_2(a_2)  \rho_1(y_1 | a_1, a_2) .\]

We maintain the following assumption on signals.

\begin{assumption} \label{a:signal} \hfill
\label{stage game noise assumption}
\begin{enumerate}
    \item The distribution of the signal $y_0$ has full support: $\rho_0(y_0 | a_0) > 0$ for all $y_0 \in Y_0, a_0 \in A_0$.
    \item The support of the distribution of the signal $y_1$ is independent of $a_2$: $\rho_1(y_1|a_1,a_2)>0 \implies \rho_1(y_1|a_1,a'_2)>0$ for all $y_1 \in Y_1, a_1\in A_1, a_2\neq a_2'\in A_2$.\footnote{This assumption rules out perfect monitoring of player 2's action. However, this is unnecessary: the analysis is unaffected by the introduction of an arbitrary additional public signal $y_2$, independent of $(y_0,y_1)$ conditional on $(a_0,a_1,a_2)$, whose distribution $\rho_2(y_2|a_2)$ depends only on $a_2$.}
    \item The signal $y_1$ statistically identifies the long-run player's action: for any $a_2 \in A_2$, the $|A_1|$ vectors $\rho_1(\cdot | a_1,a_2)_{a_1 \in A_1}$ are linearly independent in $\bb{R}^{|Y_1|}$.
\end{enumerate}
\end{assumption}

However, we emphasize that, while the public signal $y_1$ identifies player 1's action $a_1$, it does not identify her strategy $s_1$ (whenever $|Y_0|\geq 2$), because $y_0$ is player 1's private information.

Throughout the paper, for any joint distribution $\chi \in \Delta (X_1 \times X_2)$ over a product set $X_1 \times X_2$, $\pi_{X_i}(\chi)$ denotes its marginal on $X_i$. We also denote the marginal of $\gamma(\alpha_0,s_1)$ over $Y_0$ (which depends only on $\alpha_0$) by $\rho(\alpha_0)=\pi_{Y_0} (\gamma(\alpha_0,s_1))$, and we denote its marginal over $A_1$ by $\phi(\alpha_0,s_1)=\pi_{A_1} (\gamma(\alpha_0,s_1))$.

The players' payoff functions are given by $u_0: A_0 \times A_1 \to \bb{R}$ for player $0$ and $u_i: Y_0 \times A_1 \times A_2 \to \bb{R}$ for players $i \in \{1,2\}$. Thus, player 0's payoff depends on his own action and player 1's action, while the payoffs of players 1 and 2 depend on their actions and the signal $y_0$. The assumption that player 0's payoff does not depend on player 2's action simplifies the analysis and is satisfied in our applications.\footnote{However, none of our results requires this assumption, with the exception of Theorem \ref{salience payoff theorem} in Section \ref{s:salience}.} Finally, in a slight abuse of notation, we also write $u_i(\alpha_0,s_1,\alpha_2)$ for player $i$'s expected payoff at stage-game strategy profile $(\alpha_0,s_1,\alpha_2)$, and we let $\underline{u}_1 = \min_{a_0,s_1,a_2}u_1(a_0,s_1,a_2)$ and $\bar{u}_1 = \max_{a_0,s_1,a_2}u_1(a_0,s_1,a_2)$.

Deterrence games fit this framework by making $A_2$ a singleton, which effectively drops player 2 from the model. Trust and delegation games fit by making $A_0$ a singleton (i.e., making player 0 Nature). Communication games fit by making $A_0$ a singleton; viewing $\rho_0(y_0)$ as the prior distribution of a payoff-relevant state $y_0$; letting $\rho(y_1|a_1,a_2)=\mathbf{1}(\{y_1=a_1\})$ (so $a_1$ is perfectly monitored); viewing $a_2$ as a mapping from $a_1$ to a finite set of responses $R$; and assuming that $u_1$ and $u_2$ depend on $a_2$ only through the induced response $a_2(a_1) \in R$.

We conclude this subsection by adapting some definitions from \cite{FudenbergLevine92}. For any strategy $s_1$, let $B(s_1) \subset \Delta(A_0) \times \Delta(A_2)$ be the set of short-run player strategies $(\alpha_0, \alpha_2)$ satisfying 
\begin{align*}
   \supp(\alpha_0) \subset  \argmax_{a_0 \in A_0} u_0(a_0, s_1) \qquad \text{and} \qquad 
    \supp(\alpha_2) \subset  \argmax_{a_2 \in A_2} u_2(\alpha_0, s_1, a_2) , 
\end{align*}
so that player $0$ best responds to $s_1$, and player $2$ best responds to $\alpha_0$ and $s_1$. With this notation, the long-run player's \textit{(lower) Stackelberg payoff} is
\[ v_1^* =  \sup_{s_1 \in \Delta(A_1)^{Y_0}} \inf_{ (\alpha_0, \alpha_2) \in B(s_1) } u_1(\alpha_0, s_1, \alpha_2). \]
We refer to a strategy that attains this supremum as a \textit{Stackelberg strategy.} More generally, for any strategy $s_1$, we denote the corresponding lower commitment payoff by
\[ V(s_1)=  \inf_{ (\alpha_0, \alpha_2) \in B(s_1) } u_1(\alpha_0, s_1, \alpha_2).\footnote{Recall that the \emph{lower} (resp., \emph{upper}) commitment payoff results from the \emph{least favorable} (resp., \emph{most favorable}) short-run best response.} \]

Finally, we employ the following definition\footnote{Here and throughout the paper, $||\cdot ||$ denotes the $\sup$ norm.}.
\begin{definition}
    For any long-run player strategy $s_1$ and any $\eta \geq 0$, a short-run player strategy $(\alpha_{0},\alpha_{2}) \in \Delta(A_0) \times \Delta(A_2)$ is an \textit{$\eta$-confirmed best response} to $s_1$ if there exists $s_1'$ such that 
    \begin{enumerate}
        \item $(\alpha_0,\alpha_2)\in B(s_1')$, and
        \item $||p(\alpha_0, s_1, \alpha_2) - p(\alpha_0, s_1', \alpha_2)|| \leq \eta$.  
    \end{enumerate}
     For any $s_1'$ that satisfies these conditions, we say that it \textit{$\eta$-confirms $(\alpha_0, \alpha_2)$ against $s_1$}.
\end{definition}

Let $B_{\eta}(s_1)$ be the set of $\eta$-confirmed best responses to $s_1$. Note that $B_1(s_1) \supset B_{\eta'}(s_1) \supset B_{\eta}(s_1) \supset B(s_1)$ for all $\eta' \geq \eta \geq 0$, where $B_{0}(s_1) = B(s_1)$ if $s_1$ is identified (which, again, is not the case in our model whenever $|Y_0|\geq 2$), and $B_1(s_1)$ is the set of all short-run player strategies that best respond to \textit{some} long-run player strategy. Since $B_1(s_1)$ does not depend on $s_1$, we abbreviate it to $B_1$.
In addition, $B_\eta(s_1)\downarrow B_0(s_1)$ as $\eta \to 0$, by upper hemi-continuity of the short-run players' best-response correspondences. 
Finally, for any strategy $s_1$, we denote the lower commitment payoff when the short-run players take a $0$-confirmed best response by
\[V_0(s_1)=  \inf_{ (\alpha_0, \alpha_2) \in B_0(s_1) } u_1(\alpha_0, s_1, \alpha_2).\]
Note that $V(s_1) \geq V_0(s_1)$ for each $s_1$, since $B(s_1) \subset B_0(s_1)$.

\subsection{The Repeated Game} \label{s:repeated}

The stage game is repeated in each period $t = 0, 1, 2, \dots$. Player $1$ is a long-lived player with discount factor $\delta \in (0, 1)$, while players $0$ and $2$ are short-lived and take myopic best replies after observing only the public history of signals. A period-$t$ public history is denoted $h^t=(y_{1,t'})_{t'=0}^{t-1} \in Y_1^t$. Let $H^t$ be the set of period-$t$ (public) histories, $H=\bigcup_t H^t$ the set of all finite histories, and $H^\infty=Y_1^\infty$ the set of infinite histories. A repeated game strategy $\sigma_i$ for player $i$ maps public histories to stage game strategies: formally, $\sigma_i$ is a function from $H$ to $\Delta(A_i)$ for $i \in \{0,2\}$, and is a function from $H$ to $\Delta(A_1)^{Y_0}$ for $i=1$.\footnote{In principle, the long-run player could condition on her past private signals and actions in addition to the public history, but allowing this does not affect the set of equilibrium payoffs, because short-run player strategies are measurable with respect to the public history and long-run player payoffs are independent of past signals and actions.}

The long-run player's \textit{type}, denoted $\omega \in \Omega$, is either \emph{rational} ($\omega=\omega_R$) or is one of a countable number of \emph{commitment types} indexed by a (potentially mixed) stage game strategy $s_1 \in \Delta(A_1)^{Y_0}$, where type $\omega_{s_1}$ plays $s_1$ in every period.\footnote{We thus assume that there is only one rational type (unlike \cite{FudenbergLevine92}, who allow multiple rational types). We discuss the extension to multiple rational types in Section \ref{s:discussion}.} The type $\omega$ is drawn according to a full-support prior $\mu_0 \in \Delta(\Omega)$ at the start of the game and is perfectly persistent. We study $\underline{U}_1(\delta)$ and $\bar{U}_1(\delta)$, the infimum and supremum of the (rational) long-run player's payoff in any Nash equilibrium $(\sigma_0^*, \sigma_1^*, \sigma_2^*)$ of this incomplete-information repeated game. Here and throughout the paper, $\sigma_1^*$ denotes the equilibrium strategy for the rational long-run player, and we let $\bar \sigma_1^*$ denote the corresponding unconditional long-run player strategy, averaging over $\Omega$ by updating the prior $\mu_0$ by Bayes' rule. Note that $\bar \sigma_1^*(h^t)$ is defined only for histories $h^t$ that arise on path for some long-run player strategy. Finally, given a strategy profile $(\sigma_0^*, \sigma_1^*, \sigma_2^*)$, we let $\bb{P} \in \Delta(H^\infty)$ denote the induced measure over infinite histories conditional on $\omega = \omega_R$, we let $\overline{\bb{P}}$ denote the corresponding unconditional measure (averaging over $\Omega$), and we let $\bb{P}_t$ and $\overline{\bb{P}}_t$ denote the corresponding projections on $H^t$. Note that the set of on-path period-$t$ histories under rational play is $\supp(\bb{P}_t)$, and that the set of period-$t$ histories where $\bar \sigma_1^*(h^t)$ is well-defined is the larger set $\supp(\overline{\bb{P}}_t)\supset \supp(\bb{P}_t)$.

The key prior result in this context is the following.\footnote{Technically, our model is not a special case of \citeauthor{FudenbergLevine92}'s because they assume a single short-run opponent in each period, but the same argument applies. Other than this difference, our model specializes theirs to the case where the long-run player's strategy is a function $s_1:Y_0 \to \Delta(A_1)$ and Assumption \ref{a:signal} holds.}

\begin{theorem0}[\citeauthor{FudenbergLevine92}, \citeyear{FudenbergLevine92}]
\label{theorem0: fl92}
For any strategy $s_1^*$, if $\omega_{s_1^*} \in \Omega$ then
\[\liminf_{\delta \to 1} \underline{U}_1 (\delta) \geq V_0(s_1^*).\]
\end{theorem0}
\label{theorem0}

Our main contribution is providing conditions on $s_1^*$ under which this bound can be improved to $V(s_1^*)$. As we saw in Section \ref{s:deterrence}, this improvement can mean the difference between securing the minmax payoff and the Stackelberg payoff.

\section{The Commitment Payoff Theorem} \label{s:general}

This section presents our main result: a patient long-run player can secure the commitment payoff $V(s_1^*)$ corresponding to any strategy $s_1^*$ such that $\omega_{s_1^*} \in \Omega$ and $s_1^*$ is confounding-defeating and not behaviorally confounded.

\subsection{The Confound-Defeating Property}

We give two equivalent definitions of the confound-defeating property. The first definition is more useful for proving our main result. The second definition is more elegant and is easier to characterize in applications (as we do in Sections \ref{s:cyclical}--\ref{s:commmunication}). The second definition is stated in optimal transport terms: for any two distributions $\rho \in \Delta (Y_0)$ and $\phi \in \Delta (A_1)$, and any strategy for player 2 $\alpha_2 \in \Delta(A_2)$, define the optimal transport problem
\[\text{OT}(\rho,\phi;\alpha_2): \quad \max_{\gamma \in \Delta(Y_0 \times A_1)} \int u_1(y_0, a_1, \alpha_2) d\gamma \quad \text{  s.t.  } \quad \pi_{Y_0}(\gamma) = \rho \text{   and  } \pi_{A_1}(\gamma) = \phi.
\]

\begin{definition}
\label{confound defeating}
     Strategy $s^*_1$ is \emph{confound-defeating} if it satisfies one of the following conditions:
     \begin{enumerate}
         \item For all $\varepsilon > 0$, there exists $\eta > 0$ such that for any $(\alpha_0, \alpha_2) \in B_\eta (s^*_1)$ and any $s_1'$ satisfying $||s_1' - s^*_1|| > \varepsilon$ but $||p(\alpha_0, s_1', \alpha_2) - p(\alpha_0, s_1^*, \alpha_2)|| < \eta$, there exists $\tilde s_1$ satisfying $p(\alpha_0, \tilde s_1, \alpha_2) = p(\alpha_0, s_1', \alpha_2)$ and $u_1(\alpha_0, \tilde s_1, \alpha_2) > u_1(\alpha_0, s_1', \alpha_2)$.

         \item For any $(\alpha_0, \alpha_2) \in B_0 (s^*_1)$, $\gamma(\alpha_0,s_1^*)$ is the unique solution to $\text{OT}(\rho(\alpha_0),\phi(\alpha_0,s_1^*);\alpha_2)$.
    \end{enumerate}
\end{definition}

The first definition says that a strategy $s_1^*$ is confound-defeating if any strategy $s_1'$ that is a possible confound---in that it differs significantly from $s_1^*$ but induces a similar marginal over signals against some $\eta$-confirmed best response---is undetectably dominated---in that the long-run player is strictly better-off under a different strategy $\tilde{s}$ that induces the same marginal. The second definition says that $s_1^*$ itself undetectably dominates any strategy $s_1'$ that induces the same marginal over signals against some $0$-confirmed best response.

\begin{proposition} \label{p:CD}
    The two definitions of confound-defeatingness are equivalent.
\end{proposition}

It is immediate that if $\gamma(\alpha_0,s_1^*)$ is not the unique solution to $\text{OT}(\rho(\alpha_0),\phi(\alpha_0,s_1^*);\alpha_2)$ then the first definition of confound-defeatingness cannot hold. The converse relies on Assumption \ref{a:signal}(3) (player 1's action is identified). Without identification, the first definition of confound-defeatingness still gives a commitment payoff theorem, but it is more difficult to check and does not reduce to monotonicity in one-dimensional supermodular games.

The following lemma gives the key implication of confound-defeatingness: in any Nash equilibrium, at any on-path history under rational play where the marginal over signals is close to that induced by a confound-defeating strategy $s_1^*$---both unconditionally and conditional on the event the long-run player is rational---the rational long-run player must play a strategy close to $s^*_1$. Here and throughout the paper, given a repeated game strategy profile $(\sigma_0,\sigma_1,\sigma_2)$ and a period-$t$ history $h^t$, we abbreviate $p(\sigma_0(h^t),\sigma_1(h^t),\sigma_2(h^t))$ to $p(\sigma_0,\sigma_1,\sigma_2|h^t)$.

\begin{lemma}
    \label{confound defeating payoff}
   Fix a Nash equilibrium $(\sigma_0^*, \sigma_1^*, \sigma_2^*)$ and suppose that $s_1^*$ is confound-defeating. 
   Then for all $\varepsilon > 0$, there exists $\eta > 0$ such that, for any history $h^t \in \mathrm{supp}(\bb{P}_t)$ where 
   \begin{enumerate}
       \item $||p(\sigma_0^*, \bar \sigma_1^*, \sigma_2^*|h^t) - p(\sigma_0^*, s_1^*, \sigma_2^*|h^t)|| < \eta$, and
       \item $||p(\sigma_0^*, \sigma_1^*, \sigma_2^*|h^t) - p(\sigma_0^*, s_1^*, \sigma_2^*|h^t)|| < \eta$,
   \end{enumerate}
   we have $||\sigma_1^*(h^t) - s_1^*|| \leq \varepsilon$. 
\end{lemma}
\begin{proof}
Suppose not, so there exists a history $h^t \in \mathrm{supp}(\bb{P}_t)$ where conditions (1) and (2) hold, but 
$||\sigma_1^*(h^t) - s_1^*|| > \varepsilon$. 
Condition (1) and the fact $(\sigma_0^*, \sigma_1^*, \sigma_2^*)$ is an equilibrium imply that  $(\sigma_0^*(h^t),\sigma_2^*(h^t))$ is an $\eta$-confirmed best reply to $s_1^*$, as $
\bar \sigma_1^*(h^t)$ $\eta$-confirms it against $s_1^*$. 
Hence, condition (2) along with $||\sigma_1^*(h^t) - s_1^*|| > \varepsilon$ and confound-defeatingness imply that there exists some strategy $\tilde s_1$ such that $p(\sigma_0^*, \tilde s_1, \sigma_2^*|h^t) = p(\sigma_0^*, \sigma_1^*, \sigma_2^*|h^t)$ and $u_1(\sigma_0^*, \tilde s_1, \sigma_2^*|h^t) > u_1(\sigma_0^*, \sigma_1^*, \sigma_2^*|h^t)$. But this implies that if the long-run player deviates from $\sigma_1^*(h^t)$ to $\tilde s_1$ at $h^t$, her continuation payoff is unchanged while her stage game payoff increases. So, since $h^t \in \mathrm{supp}(\bb{P}_t)$, this deviation is strictly profitable, contradicting the assumption that $(\sigma_0^*, \sigma_1^*, \sigma_2^*)$ is an equilibrium. 
\end{proof}

\subsection{Behavioral Confounding} \label{s:confounding}

Lemma \ref{confound defeating payoff} implies that if a strategy $s_1^*$ is confound-defeating, it will not be confounded by the equilibrium play of the rational type of player 1. However, there remains the possibility of confounding by behavioral types. To address this, we make use of the following definition. 

\begin{definition}
    \label{behavioral confounding}
    Strategy $s_1^*$ is \emph{not behaviorally confounded} if, for any $\omega_{s'_1} \in \Omega$ such that $s'_1 \neq s_1^*$ and any $(\alpha_0,\alpha_2) \in B_1$, we have $p(\alpha_0, s_1^*, \alpha_2) \neq p(\alpha_0, s_1', \alpha_2)$.
\end{definition}

A strategy is not behaviorally confounded if the public signal distinguishes it from any other commitment type, whenever the short-run players take actions that best respond to some long-run player strategy. The definition allows the possibility that $s_1$ is indistinguishable from a mixture of two commitment types $\omega_{s'_1},\omega_{s''_1} \in \Omega$.\footnote{Ruling out this possibility would simplify the proof of Lemma \ref{almost uniform convergence} and would also let us replace $B_1$ with $B_0(s_1)$ in Definition \ref{behavioral confounding}. In particular, all of our results go through with the alternative definition that $s_1^*$ is not behaviorally confounded if, for any $(\alpha_0,\alpha_2) \in B_0(s_1)$, $p(\alpha_0, s_1^*, \alpha_2)$ lies outside the convex hull of the set $\bigcup_{s_1'\neq s_1^*:\omega_{s_1'}\in \Omega} p(\alpha_0, s_1', \alpha_2)$.} Note also that if there is only one commitment type $\omega_{s_1}$ then $s_1$ is not behaviorally confounded.

Our main result assumes that $s_1^*$ is not behaviorally confounded. In games where player $0$ is Nature (like trust and communication games), this is fairly innocuous, as $\alpha_0$ is exogenous, so the identification condition $p(\alpha_0, s_1^*, \alpha_2) \neq p(\alpha_0, s_1', \alpha_2)$ for all $\alpha_2 \in B_1$ holds for generic $s_1^*\neq s'_1$. In games with a ``real'' player $0$, it is much more restrictive, because $\alpha_0$ is endogenous, so the identification condition need not hold generically. For example, in the deterrence game in Section \ref{s:deterrence}, the pure Stackelberg strategy $(A,F)$ is not behaviorally confounded if and only if each other type $\omega_{s'_1} \in \Omega$ satisfies either $ps'_1(A|c)+(1-p)s'_1(A|d)>p$ and $(1-p)s'_1(A|c)+ps'_1(A|d)>1-p$, or $ps'_1(A|c)+(1-p)s'_1(A|d)<p$ and $(1-p)s'_1(A|c)+ps'_1(A|d)<1-p$. Nonetheless, in Section \ref{s:salience} we show that our results extend even if $s_1^*$ is behaviorally confounded, so long as it has sufficiently high prior weight relative to any behavioral confound that induces an $\eta$-confirmed best response that is not also a best response to $s^*_1$. 

\subsection{Payoff Lower Bound} \label{s:stackelberg}

We are now prepared to state our main result. 
\begin{theorem}
\label{main payoff commitment theorem}
For any strategy $s_1^*$, if $\omega_{s^*_1} \in \Omega$ and $s_1^*$ is confound-defeating and not behaviorally confounded, then
    \[ \liminf_{\delta \to 1} \und U_1(\delta) \geq V(s^*_1). \]
\end{theorem}

In particular, if $s^*_1$ is a Stackelberg strategy, Theorem \ref{main payoff commitment theorem} implies that a patient long-run player can secure her Stackelberg payoff $v^*_1$.

The logic of Theorem \ref{main payoff commitment theorem} is as follows. Fix any equilibrium, and suppose player 1 deviates by taking $s^*_1$ in every period. By standard arguments \citep{FudenbergLevine92,sorin1999merging,Gossner11}, the short-run players eventually come to expect the signal distribution $p(\sigma^*_0,s^*_1,\sigma^*_2 | h^t)$ at public history $h^t$. Since $s^*_1$ is confound-defeating, by Lemma \ref{confound defeating payoff}, the short-run players additionally come to expect that if player 1 is rational, she plays a stage game strategy close to $s^*_1$. Since $s^*_1$ is not behaviorally confounded, the short-run players also eventually learn that player 1 is not some commitment type other than $\omega_{s^*_1}$.\footnote{This step is not trivial, for example because we allow $s^*_1$ to be indistinguishable from a mixture of commitment types.} In total, the short-run players come to believe that player 1 is either the commitment type $\omega_{s^*_1}$ or is rational and playing a stage game strategy close to $s^*_1$. This leads the short-run players to best respond to $s^*_1$, which ensures the long-run player a payoff of at least $V(s^*_1)$. 

 \begin{proof}
Fix any $\varepsilon > 0$. We show that there exists $\bar \delta < 1$ such that for all $\delta > \bar \delta$, we have $\und U_1(\delta) \geq  V(s^*_1) - \varepsilon$. To do so, we fix any Nash equilibrium $(\sigma_0^*, \sigma_1^*, \sigma_2^*)$ and show that player 1's payoff from deviating by always taking $s^*_1$ is at least $V(s^*_1) - \varepsilon$. Let $\mathbb{Q} \in \Delta(H^\infty)$ denote the probability measure over infinite histories $H^\infty$ induced by this deviation: i.e., the measure induced by strategy profile $(\sigma_0^*, s_1^*, \sigma_2^*)$. 
We proceed in four steps. 

\emph{Step 1: Teaching the Marginal Over Signals.} For any $\eta >0$, define the set of period-$t$ histories where the equilibrium signal distribution is within $\eta$ of that under the deviation by
\[H_\eta^t=\left\{ h^t : ||p(\sigma_0^*,s^*_1, \sigma_2^* | h^t)  - p(\sigma_0^*, \bar \sigma^*_1, \sigma_2^* | h^t)|| \leq \eta \right\}.\]
Note $H_\eta^t \subset \supp(\bar{\bb{P}})$ as $p(\sigma_0^*, \bar \sigma^*_1, \sigma_2^* | h^t)$ is only well-defined at these history; this is without loss of generality as $\bb{Q}, \bb{P}$ are absolutely continuous with respect to $\bar{\bb{P}}$.
We recall a standard bound (essentially due to \cite{Gossner11}) on the expected number of periods $t$ where $h^t \notin H_\eta^t$. We include a proof in Appendix \ref{gossner proof}.

\begin{lemma} \label{l:gossner}
We have
\[ \mathbb{E}^{\mathbb{Q}} \left[ \# \left\{ t: h^t \notin H_\eta^t) \right\} \right] < \bar T(\eta, \mu_0) := -\frac{2 \log \mu_0(\omega_{s_1^*})}{\eta^2}. \]
\end{lemma}

\emph{Step 2: Ruling Out ``Bad'' Commitment Types.} For any $\zeta >0$, denote the set of beliefs with at most $\zeta$ weight on commitment types other than $s^*_1$ by 
 \[M_\zeta= \left\{ \mu \in \Delta (\Omega) : \mu \left(\{ \omega_R,\omega_{s^*_{1}} \}\right) \geq 1-\zeta \right\}.\]
 The next lemma shows that beliefs under $\mathbb{Q}$ concentrate on $M_0$ with high probability, uniformly in $\delta$. The proof, which relies on the martingale convergence theorem, Assumption \ref{a:signal}(1) and \ref{a:signal}(2), and the assumption that $s_1^*$ is not behaviorally confounded, is deferred to Appendix \ref{uniform convergence proof}. In what follows, given a history $h$, $\mu_t(\cdot |h) \in \Delta(\Omega)$ denotes the posterior belief over $\Omega$ conditional on the period $t$ truncation $h^t$ of $h$. (We also write $\mu_t(\cdot |h^t)$ for the posterior belief at period-$t$ history $h^t$.)

\begin{lemma}
\label{almost uniform convergence}
For all $\zeta > 0$, there exists a set of infinite histories $G(\zeta) \subset H^\infty$ satisfying $\mathbb{Q}(G(\zeta))>1-\zeta$ and a period $\hat{T}(\zeta)$ (independent of $\delta$ and the choice of equilibrium) such that, for any $h \in G(\zeta)$ and any $t \geq \hat{T}(\zeta)$, we have $\mu_t(\cdot |h) \in M_\zeta$.
\end{lemma}

\emph{Step 3: Inducing Short-Run Best Responses.} For any $\xi >0$, we say that a short-run player strategy $(\alpha_0, \alpha_2)$ is a \emph{$\xi$-close best response to $s_1^*$} (denoted $(\alpha_0, \alpha_2) \in \hat{B}_{\xi}(s_1^*)$) if $(\alpha_0, \alpha_2) \in B(s_1)$ for some $s_1$ such that $||s_1 - s_1^*|| < \xi$. Since $\hat{B}_{\xi}(s_1^*)$ is upper hemi-continuous and $\hat{B}_{0}(s_1^*) = B(s_1^*)$, we have
\begin{align*}
   \liminf_{\xi \to 0} \inf_{(\alpha_0, \alpha_2) \in \hat{B}_{\xi}(s^*_1)} u_1(\alpha_0, s_1^*, \alpha_2) \geq \inf_{(\alpha_0, \alpha_2) \in B(s_1^*)} u_1(\alpha_0, s_1^*, \alpha_2) = V(s_1^*).
\end{align*}
The next lemma shows that if the short-run players expect the marginal induced by $s_1^*$ \emph{and} believe that player 1 is either the $s_1^*$ commitment type or the rational type, they will take a $\xi$-close best response to $s_1^*$. Its proof (deferred to Appendix \ref{almost close proof}) relies on Lemma \ref{confound defeating payoff}.

\begin{lemma}
\label{almost close}
There exist strictly positive functions $\zeta(\eta)$ and $\xi(\eta)$, satisfying $\lim_{\eta \to 0}\zeta(\eta)=\lim_{\eta \to 0}\xi(\eta)=0$, such that if $h^t \in H_\eta^t$ and $\mu_t (\cdot|h^t) \in M_{\zeta(\eta)}$ then $(\sigma^*_0(h^t),\sigma^*_2(h^t)) \in \hat{B}_{\xi(\eta)}(s^*_1)$.
\end{lemma}

\emph{Step 4: Completing the Proof.} By Lemmas \ref{l:gossner} and \ref{almost uniform convergence}, conditional on the (at least) probability $1-\zeta(\eta)$ event that $h \in G(\zeta(\eta))$, the expected number of periods where either $h^t \notin H_\eta^t$ or $\mu_t (\cdot|h^t) \notin M_{\zeta(\eta)}$ is at most $\bar{T}(\eta,\mu_0)+\hat{T}(\zeta(\eta))$. By Lemma \ref{almost close}, in any period where $h^t \in H_\eta^t$ and $\mu_t (\cdot|h^t) \in M_{\zeta(\eta)}$, we have $(\sigma^*_0(h^t),\sigma^*_2(h^t)) \in \hat{B}_{\xi(\eta)}(s^*_1)$, and hence, for any sufficiently small $\eta$,
\begin{align*}
    u_1(\sigma_0^*(h^t), s_1^*, \sigma_2^*(h^t)) \geq \liminf_{\eta \to 0} \inf_{(\alpha_0, \alpha_2) \in \hat{B}_{\xi(\eta)}(s^*_1)} u_1(\alpha_0, s_1^*, \alpha_2) - \frac{\varepsilon}{3} \geq V(s_1^*) - \frac{\varepsilon}{3}.
\end{align*}

Front-loading the expected periods where $h^t \notin H_\eta^t$ or $\mu_t (\cdot|h^t) \notin M_{\zeta(\eta)}$ (and positing the minimum payoff of $\underline{u}_1$ in these periods) gives a lower bound for player 1's payoff of
\begin{align*}
    \left(1 - \delta^{\bar{T}(\eta,\mu_0)+\hat{T}(\zeta(\eta))}\right)\underline{u}_1 + \delta^{\bar{T}(\eta,\mu_0)+\hat{T}(\zeta(\eta))}\left(  (1 - \zeta(\eta)) \bar V(s_1^*)  + \zeta(\eta) \underline{u}_1  - \frac{\varepsilon}{3} \right) 
\end{align*}
As $\delta \to 1$, this lower bound converges to 
\[ (1 - \zeta(\eta)) \bar V(s_1^*)  + \zeta(\eta) \underline{u}_1  - \frac{\varepsilon}{3}. \]
By continuity, at a cost of $\varepsilon/3$, this bound remains valid for all large enough $\delta < 1$. Finally, taking $\eta$ to be small enough so $\zeta(\eta)(\bar V(s_1^*)-\underline{u}_1) <\varepsilon/3$ gives the desired bound of $V(s_1^*) - \varepsilon$. 
\end{proof}

\section{Cyclical Monotonicity and Payoff Upper Bound} \label{s:cyclical}

This section characterizes confound-defeating strategies in terms of the support of the joint distributions over $Y_0 \times A_1$ they induce, and uses this characterization to develop a partial converse to Theorem \ref{main payoff commitment theorem}.

\subsection{Cyclical Monotonicity}

Our characterization is based on the following strict version of the familiar notion of cyclical monotonicity \citep{rochet1987necessary}. The definition and subsequent characterization are elementary, but we are not aware of a reference.\footnote{The closest argument we are aware of is the proof of Lemma 2 of \cite{ball2024quota}. We thank Ian Ball for pointing out this connection and suggesting the proof of Proposition \ref{p:cyclical}.}

\begin{definition} \label{d:CM}
     Fix finite sets $X,Y$ and a function $u:X\times Y \to \bb{R}$. A set $S \subset X \times Y$ is \emph{strictly $u$-cyclically monotone} if for any finite collection of pairs $\{(x_i, y_i)_{i = 1}^N\} \subset S$ such that $\left\{(x_i, y_i)_{i=1}^N\right\} \neq \left\{(x_i, y_{i+1})_{i=1}^N\right\}$ (with convention $y_{N+1}=y_1$), 
    \[ \sum_{i = 1}^N u(x_i, y_i) > \sum_{i = 1}^N u(x_i, y_{i+1}). \]
\end{definition}

\begin{proposition} \label{p:cyclical}
    A joint distribution $\gamma \in \Delta (X \times Y)$ satisfying $\pi_{X}(\gamma) = \rho$ and $ \pi_{Y}(\gamma) = \phi$ is the unique solution to the optimal transport problem
    \[\text{OT}(\rho,\phi): \quad \max_{\gamma' \in \Delta(X \times Y)} \int u(x,y) d\gamma' \quad \text{such that} \quad \pi_{X}(\gamma') = \rho \text{   and  } \pi_{Y}(\gamma') = \phi
\] if and only if its support $\supp(\gamma) \subset X \times Y$ is strictly $u$-cyclically monotone.
\end{proposition}

We apply Proposition \ref{p:cyclical} to our setting with $X=Y_0$ and $Y=A_1$ and use the OT definition of confound-defeatingness to characterize confound-defeatingness in terms of the support of $\gamma(\alpha_0,s_1^*)$. Since we have assumed that the distribution of $y_0$ has full support, this set depends only on $s_1^*$, so we write it as \[\supp(s_1^*):=\left\{ (y_0,a_1)\in Y_0 \times A_1 \; : \: a_1\in \supp(s_1^*(y_0)) \right\}.\]
In addition, letting $u_1(\cdot,\alpha_2)$ denote player 1's utility $u_1(y_0,a_1,\alpha_2)$ as a function of $(y_0,a_1)$ for a fixed player 2 strategy $\alpha_2$, we say that $u_1$ is \emph{strictly cyclically separable} if whenever a set $S \subset Y_0 \times A_1$ is strictly $u_1(\cdot,\alpha_2)$-cyclically monotone for \emph{some} $\alpha_2$, it is strictly $u_1(\cdot,\alpha_2)$-cyclically monotone for \emph{all} $\alpha_2$. In this case, the strict $u_1$-cyclical monotonicity of a set $S \subset Y_0 \times A_1$ is well-defined independent of $\alpha_2$.\footnote{Note that this case always applies in games without a player 2, such as deterrence games.} Finally, we say that a strategy $s_1^*$ is \emph{strictly $u_1(\cdot,\alpha_2)$ (resp., $u_1$)-cyclically monotone} if $\supp(s_1^*)\subset Y_0 \times A_1$ is strictly $u_1(\cdot,\alpha_2)$ (resp., $u_1$)-cyclically monotone. We obtain the following characterization.

\begin{corollary}\label{c:CM}
    A strategy $s_1^*$ is confound-defeating if and only if it is strictly $u_1(\cdot,\alpha_2)$-cyclically monotone for all $(\alpha_0,\alpha_2) \in B_0(s_1^*)$.

    Moreover, if $u_1$ is strictly cyclically separable, a strategy $s_1^*$ is confound-defeating if and only if it is strictly $u_1$-cyclically monotone.
\end{corollary}

Together with Theorem \ref{main payoff commitment theorem}, we obtain the following corollary, where we denote the long-run player's greatest lower commitment payoff from a strictly $u_1$-cyclically monotone strategy by
\[\underline{v}_1^{\mathrm{CM}}:=\sup_{s_1 :\substack{ \text{ $\omega_{s_1} \in \Omega$ and $s_1$ is strictly $u_1$-cyclically}
   \\ \text{ monotone and not behaviorally confounded}}} \min_{(\alpha_0, \alpha_2) \in B(s_1)} u_1(\alpha_0, s_1, \alpha_2).\]

\begin{corollary}\label{c:commitment payoff}
Suppose $u_1$ is strictly cyclically separable. For any strategy $s_1^*$, if $\omega_{s_1^*}\in \Omega$, and $s_1^*$ is strictly $u_1$-cyclically monotone and not behaviorally confounded, then
    \[ \liminf_{\delta \to 1} \und U_1(\delta) \geq V(s_1^*). \] 
    In particular, 
    \[ \liminf_{\delta \to 1} \und U_1(\delta) \geq \underline{v}_1^{\mathrm{CM}}. \]
    \end{corollary}

Thus, when the long-run player's payoff is strictly cyclically separable, the confound-defeating property in Theorem \ref{main payoff commitment theorem} can be replaced by strict $u_1$-cyclical monotonicity.

\subsection{Payoff Upper Bound} \label{s:converse}

We now give a partial converse to Corollary \ref{c:commitment payoff}: if the long-run player is rational with high probability, her payoff is bounded above by her greatest upper commitment payoff from any $u_1$-cyclically monotone strategy, defined by
\[\bar v_1^{\mathrm{CM}}:=\sup_{s_1 \; : \text{ $s_1$ is $u_1$-cyclically monotone}} \max_{(\alpha_0, \alpha_2) \in B(s_1)} u_1(\alpha_0, s_1, \alpha_2),\]
where $s_1$ is \emph{$u_1$-cyclically monotone} if $\supp(s_1)$ satisfies the usual definition of $u_1$-cyclical monotonicity for any $\alpha_2$ (i.e., Definition \ref{d:CM} with the strict inequality replaced by a weak one). The idea is that if a strategy $s_1$ is not $u_1$-cyclically monotone strategy, then the rational long-run player has a profitable and undetectable deviation from $s_1$, so the short-run players cannot expect to face a strategy close to $s_1$ with high probability when the long-run player is rational with high probability.

In the following statement, $u_1$ is \emph{cyclically separable} if whenever a set $S \subset Y_0 \times A_1$ is $u_1(\cdot,\alpha_2)$-cyclically monotone for some $\alpha_2$, it is $u_1(\cdot,\alpha_2)$-cyclically monotone for all $\alpha_2$.

\begin{proposition}\label{p:converse}
   Suppose $u_1$ is cyclically separable. Then for all $\varepsilon > 0$, there exists $\kappa > 0$ such that, for any prior $\mu_0$ satisfying $\mu_0(\omega_R) > 1 - \kappa$ and any $\delta < 1$, 
    \[ \bar U_1(\delta) < \bar v_1^{\mathrm{CM}} + \varepsilon.\]
\end{proposition}

A key step in the proof of Proposition \ref{p:converse} is the following lemma, which says that a rational long-run player with a cyclically separable utility must play a cyclically monotone stage game strategy at every history in any repeated game Nash equilibrium. We record the lemma and its (short) proof, as it applies equally to any repeated game, with or without multiple long-run players and incomplete information.

\begin{lemma}
\label{must be monotone}
    For any Nash equilibrium $(\sigma_0^*,\sigma_1^*,\sigma_2^*)$ and any history $h^t \in \mathrm{supp}(\bb{P}_t)$, $\sigma_1^*(h^t)$ is $u_1$-cyclically monotone. 
\end{lemma}

\begin{proof}
Note that $\sigma_1^*(h^t)$ must solve
 \[ \text{OT}(\sigma_0^*(h^t),\phi(\sigma_0^*(h^t),\sigma_1^*(h^t)),\sigma_2^*(h^t)).\] This holds because, otherwise, there exists a strategy $s_1$ that gives player 1 a strictly higher payoff at history $h^t$ than $\sigma_1^*(h^t)$ does, but gives the same signal distribution, and hence the same continuation payoff. By a standard optimal transport result (e.g., Theorem 1.38 in \citeauthor{santambrogio2015optimal}, \citeyear{santambrogio2015optimal}), this implies that $\supp(\gamma(\sigma_0^*(h^t), \sigma_1^*(h^t)))$ is $u_1(\cdot,\sigma_2^*(h^t))$-cyclically monotone. Hence, by cyclical separability, $\sigma_1^*(h^t)$ is $u_1$-cyclically monotone. 
\end{proof}

Given Lemma \ref{must be monotone}, at any history $h^t$ where the long-run player is thought to be rational with high probability, the short-run players must expect the long-run player strategy $\bar \sigma_1^*(h^t)$ to be $u_1$-cyclically monotone with high probability. When the short-run players best-respond to such a strategy, the long-run player's payoff cannot greatly exceed $\bar v_1^{\mathrm{CM}}$. Finally, if the prior probability of the rational type is close to $1$, then for each period $t$, the probability that $\mu_t(\omega_R|h^t ,\omega_R)$ is high is also close to $1$ (since $\mu_t(\omega_R|h^t, \omega_R)$ is a $\mathbb{P}$-submartingale), so the long-run player's overall expected payoff also cannot greatly exceed $\bar v_1^{\mathrm{CM}}$.

Combining Corollary \ref{c:commitment payoff} and Proposition \ref{p:converse} gives a fairly tight characterization of a patient long-run player's payoff when $u_1$ is cyclically and strictly cyclically separable: it is at least her lower commitment payoff from any non-behaviorally confounded, strictly $u_1$-cyclically monotone commitment type strategy, and at most her greatest upper commitment payoff from any $u_1$-cyclically monotone strategy.

\begin{remark}
    A similar argument shows that, in both our general model (without cyclical separability) and in \cite{FudenbergLevine92}'s general model, the upper Stackelberg payoff \linebreak $\sup_{s_1} \sup_{(\alpha_0,\alpha_2) \in B(s_1)} u_1(\alpha_0,s_1,\alpha_2)$ is an upper bound for $\limsup_{\mu_0:\mu_0(\omega_r)\to1} \bar U_1(\delta)$ for any $\delta<1$. The logic is that the long-run player's payoff cannot exceed the upper Stackelberg payoff at any history $h^t$ where she is thought to be rational with high probability, and such histories occur with high probability on path when $\mu_0(\omega_r)$ is close to $1$ and the long-run player is indeed rational. This is a very different upper bound than that given by \citeauthor{FudenbergLevine92}, who show that the upper Stackelberg payoff allowing $0$-confirmed best responses, $\sup_{s_1} \sup_{(\alpha_0,\alpha_2) \in B_0(s_1)} u_1(\alpha_0,s_1,s_2)$, is an upper bound for $\limsup_{\delta \to1} \bar U_1(\delta)$ for any prior $\mu_0$.
\end{remark}

\section{One-Dimensional Supermodular Games} \label{s:monotone}

\subsection{Supermodularity and Monotonicity}

We now show how our results apply in games where the long-run player's payoff is supermodular in a one-dimensional signal and action. This simple setting includes a wide range of applications, including games of deterrence, trust, delegation, and signaling.

The relevant supermodularity notion is strict supermodularity in $(y_0,a_1)$ for all $\alpha_2$.

\begin{definition}
\label{supermodularity assumption}
The long-run player's payoff $u_1$ is \textit{strictly supermodular} if there exist total orders $(\succsim_{Y_0}, \succsim_{A_1})$ such that 
\[u_1(y_0,a_1,a_2)-u_1(y_0,a'_1,a_2)>u_1(y'_0,a_1,a_2)-u_1(y'_0,a'_1,a_2) \quad \text{ for all } \quad y_0 \succ y'_0, a_1 \succ a'_1, a_2.
\]   
\end{definition}

We say that a (possibly mixed) strategy $s_1: Y_0 \to \Delta(A_1)$ is monotone if any selection from the graph of its support is monotone. Formally,

\begin{definition}
\label{monotone strategies definition}
A strategy $s_1$ is \textit{monotone} if, for any $y_0 \succ y_0'$, $a_1 \in \supp(s_1(y_0))$, and $a_1' \in \supp(s_1(y_0'))$, we have $a_1 \succsim a_1'$.
\end{definition}

The following is the key result of this section.

\begin{proposition} \label{p:monotone}
   Let $u_1$ be strictly supermodular. For any $s_1^*$, the following are equivalent:
    \begin{enumerate}
        \item $s_1^*$ is confound-defeating.
        \item $s_1^*$ is monotone.
        \item $s_1^*$ is $u_1$-cyclically monotone.
    \end{enumerate}
\end{proposition}

\begin{proof}
    
The key step is the following standard result from optimal transport.

\begin{lemma}
    \label{lemma OT}
        Suppose $u_1$ is strictly supermodular. Then, for any $(\alpha_0,\alpha_2)$, $s_1^*$ is monotone if and only if $\gamma(\alpha_0,s_1^*)$ is the unique solution to $\text{OT}(\rho(\alpha_0),\phi(\alpha_0,s_1^*);\alpha_2)$.
    \end{lemma}

\begin{proof}
By Lemma 2.8 in \citet{santambrogio2015optimal}, if $s_1^*$ is monotone then $\gamma(\alpha_0,s_1^*)$ is the unique \emph{co-monotone transport plan} between $\rho(\alpha_0)$ and $\phi(\alpha_0,s_1^*)$---that is, the unique joint distribution $\gamma \in \Delta (Y_0 \times A_1)$ with marginals $\rho(\alpha_0)$ and $\phi(\alpha_0,s_1^*)$ such that, according to $\gamma$, $y_0$ and $a_1$ are co-monotone random variables. Conversely, since $\rho(\alpha_0)$ has full support, if $s_1^*$ is not monotone then $\gamma(\alpha_0,s_1^*)$ is not co-monotone. Finally, by Theorem 2.9 and Exercise 10 in \citet{santambrogio2015optimal}, when $u_1$ is strictly supermodular, the co-monotone transport plan is the unique solution to $\text{OT}(\rho(\alpha_0),\phi(\alpha_0,s_1^*);\alpha_2)$. 
\end{proof}

By Lemma \ref{lemma OT}, if $s_1^*$ is confound-defeating then $\gamma(\alpha_0, s_1^*)$ is the unique solution to 
\\ $\text{OT}(\rho(\alpha_0),\phi(\alpha_0,s_1^*);\alpha_2)$ for any $(\alpha_0, \alpha_2) \in B_0(s_1^*)$, and hence is monotone; and, conversely, if $s_1^*$ is monotone then it is the unique solution to $\text{OT}(\rho(\alpha_0),\phi(\alpha_0,s_1^*);\alpha_2)$ for any $(\alpha_0, \alpha_2)$, and hence is confound-defeating. Moreover, $s_1^*$ is monotone if and only if $\gamma(\alpha_0,s_1^*)$ is co-monotone, as shown in the proof of Lemma \ref{lemma OT}, and $\gamma(\alpha_0,s_1^*)$ is co-monotone if and only if it is $u_1$-cyclically monotone, by Lemma 1 of \citet{lin2024credible} (see also Proposition 1 of \citet{rochet1987necessary}). This establishes the desired three-way equivalence.
\end{proof}

We now denote the long-run player's lower commitment payoff from any non-behaviorally confounded, monotone commitment type strategy and her upper commitment payoff from any monotone strategy, respectively, by
\begin{align*}
   \underline{v}_1^{\mathrm{mon}} &=\sup_{s_1 :\substack{ \text{ $\omega_{s_1} \in \Omega$ and $s_1$ is monotone}
   \\ \text{ and not behaviorally confounded}}} \min_{(\alpha_0, \alpha_2) \in B(s_1)} u_1(\alpha_0, s_1, \alpha_2), \\
   \bar v_1^{\mathrm{mon}} &=\sup_{s_1 : \text{ $s_1$ is monotone}} \max_{(\alpha_0, \alpha_2) \in B(s_1)} u_1(\alpha_0, s_1, \alpha_2).
\end{align*}
Combining Theorem \ref{main payoff commitment theorem} and Propositions \ref{p:converse} and \ref{p:monotone}, we obtain the following corollary, which specializes Corollary \ref{c:commitment payoff} and Proposition \ref{p:converse} to one-dimensional supermodular games.

\begin{corollary} \label{c:monotone}
Suppose $u_1$ is strictly supermodular. Then
    \[ \liminf_{\delta \to 1} \und U_1(\delta) \geq \underline{v}_1^{\mathrm{mon}}. \]
    Conversely, for all $\varepsilon>0$, there exists $\kappa >0$ such that for any prior $\mu_0$ satisfying $\mu_0(\omega_R)>1-\kappa$ and any $\delta <1$,
     \[\bar U_1(\delta) < \bar v_1^{\mathrm{mon}}+\varepsilon. \]
    \end{corollary}

Corollary \ref{c:monotone} is a main conclusion of this paper: in a one-dimensional supermodular game, a patient long-run player can secure her commitment payoff from any monotone, non-behaviorally confounded strategy $s_1$ such that $\mu_0(\omega_{s_1})>0$. This implies the supermodular cases of Proposition \ref{p:deterrence} (as in the deterrence game $u_1$ is strictly supermodular with the order $A \succ F$ and $C \succ D$ when $x+y<1$) and Proposition \ref{p:trust} (as in the trust game $u_1$ is strictly supermodular with the order $H \succ L$ and $T \succ N$ when $\min\{w,z\}>0$), as well as our results for delegation and communication games in the subsequent sections.

At the same time, the converse direction of Corollary \ref{c:monotone} implies the submodular cases of Propositions \ref{p:deterrence} and \ref{p:trust}. For example, in the deterrence game, if $x+y>1$ (the ``submodular case'') then $u_1$ is strictly supermodular with the order $F \succ A$ and $C \succ D$. Since any strategy that is monotone with this order takes $F$ with higher probability after $c$, the unique short-run player best response to any such strategy is $D$, implying that $\bar v_1^{\mathrm{mon}}=1-p+py$. The argument for the submodular case of Proposition \ref{p:trust} is similar.

Finally, in games where $\und v_1^{mon} = \bar v_1^{mon}$ (e.g. if short-run players have a unique best response to any monotone strategy $s_1$ and the best monotone commitment strategy $s_1^*$ satisfies $\omega_{s_1^*}\in\Omega$ and is not behaviorally confounded), Corollary \ref{c:monotone} gives a unique payoff prediction as $\mu_0(\omega_R)$ and $\delta$ both approach $1$. This holds, for example, in the deterrence and trust games. 

\subsection{Delegation Games} \label{s:delegation}

We apply Corollary \ref{c:monotone} to repeated delegation games, where in each period a short-run player 2 chooses whether or not to delegate a decision to the long-run player 1, who then takes an action $a \in A \subset \mathbb{R}$ after observing a state $\theta \in \Theta \subset \mathbb{R}$, which is drawn i.i.d.\ across periods from a full-support distribution $\rho_0 \in \Delta(\Theta)$.\footnote{\cite{lipnowski2020repeated} study repeated delegation with two long-run players and no commitment types. We thank Navin Kartik for suggesting we discuss repeated delegation games.} We assume $A$ and $\Theta$ are finite.

Formally, player 2 can choose either a safe action $a_2=S\text(afe)$, in which case the decision is delegated with probability $\varepsilon>0$ (perhaps reflecting either a tremble or the occasional necessity of delegation), or action $a_2=D\text(elegate)$, in which case delegation occurs with probability $1$. Assume that player 2's action is observed, and player 1's action is observed if and only if the decision is delegated. (The state is never observed.) The assumption that delegation always occurs with positive probability ensures that Assumption \ref{a:signal} holds. In particular, player 1's action is identified even if player 2 always takes the safe action.

Normalize the players' no-delegation payoffs to $0$, and denote a player $i$'s payoff when the decision is delegated and action $a$ is taken in state $\theta$ by $u_i(a, \theta)$. Assume that $u_1(a, \theta)$ is strictly supermodular. Then, for any long-run player strategy $s_1(\theta)$, payoffs are given by 
\[ u_i(s_1, S) = \varepsilon \bb{E}_{\theta}[u_i(s_1(\theta), \theta)] \qquad \text{  and  } \qquad u_i(s_1, D) = \bb{E}_{\theta}[u_i(s_1(\theta), \theta)].\]
Note that $u_i(s_1, D)>u_i(s_1, S)$ if and only if $u_i(s_1, D)>0.$

The following result is an immediate application of Corollary \ref{c:monotone}.

\begin{proposition} \label{p:delegation}
In repeated delegation, for any monotone strategy $s_1^*$, if $\omega_{s_1^*} \in \Omega$, $s_1^*$ is not behaviorally confounded, and $u_2(s_1^*, D) > 0$, then
$\liminf_{\delta \to 1} \underline{U}_1(\delta) \geq u_1(s_1^*, D)$.
\end{proposition}

Proposition \ref{p:delegation} contrasts with the results of \cite{ElyValimaki03} and \cite{ely2008reputation}, who provide conditions under which a patient long-run player's payoff approaches her minmax payoff in all equilibria of a repeated delegation game with commitment types. A key difference is our assumption that delegation always occurs with positive probability, so the long-run player's action is always identified. With this assumption, our Assumption \ref{a:signal} would also hold in \citeauthor{ElyValimaki03}'s model, and the commitment payoff theorem would apply whenever the Stackelberg type is present with positive probability and is not behaviorally confounded.

\section{Communication Games} \label{s:commmunication}

We now consider implications of our results for repeated communication games. Recall that the model covers communication games by viewing $\rho_0(y_0)$ as the prior distribution of a payoff-relevant state $y_0$; letting $\rho_1(y_1|a_1)=\mathbf{1}(\{y_1=a_1\})$; viewing $a_2$ as a mapping from $a_1$ to a finite set of responses $R$; and assuming that $u_1$ and $u_2$ depend on $a_2$ only through the induced response $a_2(a_1)\in R$. In this section, we refer to player 1 as the sender and player 2 as the receiver, and we relabel $y_0$ as $\theta$.

\subsection{Signaling} \label{s:signaling}

In a repeated signaling game, the state $\theta \in \Theta \subset \mathbb{R}$ is drawn i.i.d. across periods. In each period, the long-run player observes $\theta$ before taking an action $a_1 \in A_1 \subset \mathbb{R}$. The short-run player observes the current action $a_1$ (but not $\theta$) and the history of past actions (but not past states) and then takes an action $r \in R$ in response. Assume $\Theta$, $A_1$, and $R$ are finite, the long-run player's payoff is given by $v(a_1,r)-w(a_1,\theta)$ for some functions $v$ and $w$, and $w$ is strictly submodular. Thus, the long-run player's preferences over the short-run player's action $r$ are independent of the state $\theta$. Note that submodularity of $w$ is a standard assumption in signaling theory: for example, in \cite{spence1973job}, $v(a_1,r)=r$ and $w(a_1,\theta)=a_1/\theta$.

The following result is another implication of Corollary \ref{c:monotone}.

 \begin{proposition} \label{p:signaling}
  In repeated signaling, for any monotone strategy $s_1^*$, if $\omega_{s_1^*} \in \Omega$ and $s_1^*$ is not behaviorally confounded, then
$\liminf_{\delta \to 1} \underline{U}_1(\delta) \geq V(s_1^*)$.
\end{proposition}

\begin{proof}
    We have $u_1(\theta,a_1,a_2)=(1-\lambda)v(a_1,a_2(a_1))-\lambda w(a_1,\theta)$, which is strictly supermodular in $(\theta,a_1)$ because $w$ is strictly submodular. The result thus follows from Corollary \ref{c:monotone}.
\end{proof}

Thus, a patient sender with state-independent preferences over the receiver's action and a strictly submodular signaling cost can secure her best commitment payoff from any monotone signaling strategy, even if the history of states is never observed.

\subsection{Perturbed Cheap Talk} \label{s:signaling}

We now turn to the following question. Consider a cheap talk game with a state-independent utility $v(r)$ for the sender and a utility $u_2(\theta,r)$ for the receiver, so the sender's action $a_1$ is payoff irrelevant. The sender's utility $u_1(\theta,a_1,r)$ is then independent of $a_1$ and hence cannot be strictly supermodular in $(\theta, r)$. However, suppose that the sender has a ``grain of commitment power,'' in that she can publicly adjust her preferences at the beginning of the game by committing to pay a small communication cost $w(a_1,\theta)$ whenever she takes action $a_1$ in state $\theta$. We ask what commitment payoffs $V(s_1)$ can be secured by leveraging such a grain of commitment.

By the revelation principle, for any set of sender actions $A_1$ and any strategy $\hat s_1: \Theta \to \Delta(A_1)$, there exists a direct \emph{communication mechanism} $s_1:\Theta \to \Delta(R)$ such that $V(s_1)=V(\hat{s}_1)$. We restrict attention to direct communication mechanisms in this section, and in particular assume that $A_1=R$, so the sender's message $a_1=\tilde r$ can be interpreted as a recommended action for the receiver. The following result (which is an immediate consequence of Corollary \ref{c:monotone}) provides a general sufficient condition for the commitment payoff $V(s_1)$ from a communication mechanism $s_1:\Theta \to \Delta(R)$ to be approximately securable.

\begin{proposition} \label{p:persuasion}
    If a communication mechanism $s_1:\Theta \to \Delta(R)$ is monotone with respect to some order $(\succsim_\Theta,\succsim_R)$ and is such that $\omega_{s_1} \in \Omega$ and $s_1$ is not behaviorally confounded, then for any $\varepsilon>0$ and any strictly submodular cost function $w:R \times \Theta \to [0,1]$,
    \[\liminf_{\delta \to 1} \underline{U}^w(\delta) \geq (1-\varepsilon)V(s_1)-\varepsilon,\]
    where $\underline{U}^w(\delta)$ is the infimum of the long-run player's payoff in any Nash equilibrium in the repeated game where her utility is given by
    \[u_1(\theta,\tilde r,r)=(1-\varepsilon)v(r)-\varepsilon w(\tilde r,\theta).\]
\end{proposition}

An interpretation of the communication cost $w(\tilde r,\theta)$ is that this represents a ``lying cost'' \citep{chen2008selecting,kartik2009strategic} incurred by a sender who recommends action $\tilde r$ in state $\theta$. In particular, if $R=\Theta$ and the receiver's optimal action in state $\theta$ is $r=\theta$, we can interpret the sender's message $\tilde r \in \Theta$ as a report of the state, and we can interpret $w(\tilde r,\theta)$ as the lying cost associated with misreporting state $\theta$ as $\tilde r$. This example matches the main example in \cite{kartik2009strategic}, where it is likewise assumed that the lying cost $w(\tilde r,\theta)$ is strictly submodular. Proposition \ref{p:persuasion} thus implies that augmenting repeated cheap talk with a small lying cost provides a reputational foundation for any communication mechanism that is monotone with respect to some order over states and actions.

For a concrete example, consider \citeauthor{KamenicaGentzkow11}'s prosecutor-judge example, where a prosecutor discloses information about a defendant's guilt to maximize the probability of conviction. The prosecutor has state-independent utility $v(r)=1(\{r=\text{Convict}\})$. If the prosecutor's payoff is augmented with an $\varepsilon$ cost of recommending that an innocent defendant be convicted, then $w(\tilde r,\theta)$ is strictly submodular in the order  $\text{Convict} \succ \text{Acquit}$, $\text{Guilty} \succ \text{Innocent}$, and our results imply that a patient prosecutor can secure her commitment payoff.\footnote{Another implication of our results is that adding a Stackelberg commitment type and a small lying cost in the infinitely-repeated ``political correctness'' game of \cite{morris2001political} suffices to secure the Stackelberg payoff, in contrast to \citeauthor{morris2001political}'s selection of the babbling equilibrium.}

To fully operationalize Proposition \ref{p:persuasion}, it remains to characterize what mechanisms $s_1:\Theta \to \Delta(R)$ are monotone with respect to some order $(\succsim_\Theta,\succsim_R)$. To do so, let $G(s_1)$ be the bipartite graph with vertices $\Theta$ and $R$, where a state $\theta$ and an action $r$ are linked if $r \in \supp(s_1(\theta))$. We will see that if $s_1$ is monotone then $G(s_1)$ is acyclic and also does not contain what we call a ``forbidden triple.''

\begin{definition}
    A \emph{forbidden triple} for a mechanism $s_1:\Theta \to \Delta(R)$ is either
    \begin{enumerate}
        \item A set of three distinct actions $\{r_1,r_2,r_3\}$ and four distinct state $\{\theta_1,\theta_2,\theta_3, \theta_4\}$ where $r_k \in \supp(s_1(\theta_k))$ for $k\in \{1,2,3\}$ and $\{r_1,r_2,r_3\}\subset \supp(s_1(\theta_4))$; or
        \item A set of three distinct states $\{\theta_1,\theta_2,\theta_3\}$ and four distinct actions $\{r_1,r_2,r_3,r_4\}$ where $\{r_k,r_4\} \in \supp(s_1(\theta_k))$ for $k\in \{1,2,3\}$.
    \end{enumerate}
\end{definition}

\definecolor{myblue}{RGB}{80,80,160}
\definecolor{mygreen}{RGB}{80,160,80}

\begin{figure}[h]
\definecolor{myblue}{RGB}{80,80,160}
\definecolor{mygreen}{RGB}{80,160,80}
       \centering       
\begin{tikzpicture}[thick,
  every node/.style={draw,circle},
  fsnode/.style={fill=myblue},
  ssnode/.style={fill=mygreen},
  every fit/.style={ellipse,draw,inner sep=-2pt,text width=2cm},
  ->,shorten >= 3pt,shorten <= 3pt
]

\begin{scope}[start chain=going below,node distance=7mm]
\foreach \i in {1,2,...,4}
  \node[fsnode,on chain] (f\i) [label=left: $\theta_{\i}$] {};
\end{scope}

\begin{scope}[xshift=4cm,yshift=-0.5cm,start chain=going below,node distance=7mm]
\foreach \i in {1, 2, 3}
  \node[ssnode,on chain] (s\i) [label=right: $r_{\i}$] {};
\end{scope}

\node [myblue,fit=(f1) (f4),label=above:$\Theta$] {};

\node [mygreen,fit=(s1) (s3),label=above:$R$] {};

\draw (f1) -- (s1);
\draw (f2) -- (s2);
\draw (f3) -- (s3);
\draw (f4) -- (s1);
\draw (f4) -- (s2);
\draw (f4) -- (s3);

\end{tikzpicture}
\caption{A Type (1) Forbidden Triple}
\caption*{\emph{Notes.} Monotonicity is violated for any placement of $\theta_4$ in the order $\succsim_\Theta$.}
\label{fig:type1}
\end{figure}

For example, if a Type (1) forbidden triple exists, where without loss $r_1 \prec r_2 \prec r_3$ and $\theta_1 \prec \theta_2 \prec \theta_3$, then $s_1$ cannot be monotone with respect to any order, as if $\theta_4 \prec \theta_2$ then $s_1$ is non-monotone because $r_3 \in \supp (s_1(\theta_4))$ but $r_2 \in \supp (s_1(\theta_2))$; and if $\theta_4 \succ \theta_2$ then $s_1$ is non-monotone because $r_1 \in \supp (s_1(\theta_4))$ but $r_2 \in \supp (s_1(\theta_2))$. See Figure \ref{fig:type1}.

Our final result is that, conversely, if $G(s_1)$ is acyclic and does not contain a forbidden triple, then $s_1:\Theta \to \Delta(R)$ is monotone with respect to some order.

\begin{proposition} \label{p:chain}
    A communication mechanism $s_1:\Theta \to \Delta(R)$ is monotone with respect to some order $(\succsim_\Theta,\succsim_R)$ if and only if $G(s_1)$ is acyclic and does not contain a forbidden triple.
\end{proposition}

Proposition \ref{p:chain} implies that the set of mechanisms that are monotone with respect to some order includes, for example, all partitions (i.e., deterministic mechanisms $s_1:\Theta \to R$) and all linear partitions with randomization at the endpoints. This includes the set of all monotone partitions, which are shown to be optimal in certain persuasion problems by \cite{kolotilin2018optimal} and \cite{dworczak2019simple}. 
However, it does not always include the optimal mechanism. For example, if the receiver's optimal action is $r=\bb{E}[\theta]$ and $\theta \in \{0,1\}$ with equal probability, and the sender's utility is $\mathbf{1}(\{r \in \{1/3,2/3\}\})$, then the unique optimal mechanism induces $r \in \{1/3,2/3\}$ with equal probability, but this mechanism is not monotone with respect to any order because the corresponding graph $G(s_1)$ contains the cycle $(0,1/3),(1,1/3),(1,2/3),(0,2/3)$.\footnote{This example is a discrete version of a ``bi-pooling'' policy. With a continuous state, \cite{kleiner2021extreme} and \cite{arieli2023optimal} show that the bi-pooling policies are those that are uniquely optimal in some persuasion problem.}

Proposition \ref{p:chain} is a general mathematical result that could have other applications (and that may have been previously noted in other contexts, although we have not found a reference). It characterizes when a Markov transition matrix $f:X \to Y$ is consistent with a joint distribution over $X \times Y$ that is co-monotone with respect to some order $(\succsim_X,\succsim_Y)$. Alternatively, it characterizes when the vertices of a bipartite graph can be drawn on two straight lines so that no edges cross.

\section{A Payoff Bound with Behavioral Confounding}\label{s:salience}

Our last result generalizes Theorem \ref{main payoff commitment theorem} to the case where $s^*_1$ is behaviorally confounded. This extension is particularly important for games where $\alpha_0$ is endogenous, like the deterrence game in Section \ref{s:deterrence}, as in these games $s^*_1$ is often behaviorally confounded when there are multiple commitment types (as discussed in Section \ref{s:confounding}).

The details of the generalization are somewhat intricate, but the main idea is simple. If $s^*_1$ is behaviorally confounded, we calculate the minimum weight on $s^*_1$ that ensures that once short-run players learn the desired signal distribution, they best respond to $s^*_1$ (rather than a confounding strategy). Then we calculate the minimum probability $\beta$ with which the long-run weight on $s^*_1$ exceeds this level under the deviation measure $\mathbb{Q}$ (where the player 1 always plays $s_1^*$). We call $\beta$ the ``salience'' of $s^*_1$, and we establish a lower bound for the patient long-run player's payoff as a function of $\beta$. If $s^*_1$ is not behaviorally confounded then its salience is $1$, in which case we recover Theorem \ref{main payoff commitment theorem}.

Before defining salience, we require a preliminary definition. In what follows, given a belief $\mu \in \Delta(\Omega)$ and a subset $\Omega' \subset \Omega$, we denote the conditional distribution of $\mu$ over $\Omega'$ by $\mu(\cdot | \Omega')$. In addition, given a belief $\mu \in \Delta(\Omega \setminus \{\omega_R\})$, we slightly abuse notation by also denoting by $\mu(\cdot | \Omega \setminus \{\omega_R\})$ the strategy $\tilde s_1$ given by $\tilde s_1(y_0)[a_1]=\sum_{\omega_{s_1} \in \Omega \setminus \{\omega_R\}} \mu(\omega_{s_1}) s_1(y_0)[a_1]$.

\begin{definition}
\label{confounding weights}
Fix any strategy $s_1^*$ and any $\eta,\varsigma>0$, a number $c \in [0, 1]$ is a \emph{$(s_1^*,\eta, \varsigma)$-confounding weight} if there exists a belief $\mu \in \Delta(\Omega)$ satisfying the following conditions.
\begin{enumerate}
\item $\mu(\omega_R)<1$.
    \item  There exists $s_1 \in \Delta (A_1)^{Y_0}$ such that $||\mu(\cdot | \Omega \setminus \{\omega_R\}) - s_1|| < \varsigma$ and $B(s_1) \setminus B(s_1^*) \neq \emptyset$. 
    \item $\mu(\Omega_\eta (s_1^*)) > 1 -\eta$, where 
    \[\Omega_\eta (s_1^*) = \{\omega_{s_1} \in \Omega \setminus \{\omega_R\}: ||p(\alpha_0, s_1, \alpha_2) - p(\alpha_0, s_1^*, \alpha_2)|| < \eta \text{  for some  } (\alpha_0, \alpha_2) \in B_1 \} \cup \{\omega_R\}.\]
    \item $\mu(\omega_{s_1^*} | \Omega \setminus \{\omega_R\}) = c.$
\end{enumerate}
Let $c_{\eta, \varsigma}(s_1^*)$ denote the supremum over $(s_1^*,\eta,\varsigma)$-confounding weights, with the convention that if no such weight exists, then $c_{\eta, \varsigma}(s_1^*) = -\infty$. Finally, let $c_0(s_1^*) = \lim_{\varsigma \to 0} \lim_{\eta \to 0} c_{\varsigma, \eta}(s_1^*)$.
\end{definition}

Condition (1) implies $\mu(\cdot | \Omega \setminus \{\omega_R\} )$ is well-defined. Condition (2) says $\mu(\cdot | \Omega \setminus \{\omega_R\} )$ is within $\varsigma$ of a belief to which the short-run players have a best response outside $B(s_1^*)$. Condition (3) says $\mu$ puts at most $\eta$ weight on commitment types that induce signals that are not $\eta$ close to those induced by $s_1^*$. Condition (4) says $\mu(\cdot | \Omega \setminus \{\omega_R\} )$ assigns probability $c$ to $s_1^*$. Note $c_0(s_1^*)<1$ by upper hemi-continuity of the best-response correspondence $B(\cdot)$. Moreover, the sets $\Omega_\eta(s_1^*)$ are nested and all contain $s_1^*$, which implies the set 
\[ \Omega_0(s_1^*) := \A_{\eta > 0} \Omega_\eta(s_1^*) \]
is well-defined and contains $\{\omega_{s_1^*}, \omega_R\}$. 

\begin{definition} \label{d:salience}
    The \emph{salience} of a strategy $s_1^* \in \Omega$ is \[\beta = \max\left\{\frac{\mu_0(\omega_{s_1^*} | \Omega_0(s_1^*) \setminus \{\omega_R\}) - c_0(s_1^*)}{1 - c_0(s_1^*)}, 0 \right\},\] with the convention that if $c_0(s_1^*)=-\infty$ then $\beta=1$.
\end{definition}

The logic of Definitions \ref{confounding weights} and \ref{d:salience} is that conditional on the long-run player being irrational, $c_0(s_1^*)$ is the minimum weight on $s_1^*$ that ensures short-run players best respond to $s_1^*$, and (by Bayes' rule) $\beta$ is the minimum probability the long-run weight on $s_1^*$ strictly exceeds $c_0(s_1^*)$.

The general version of our main result is as follows.

\begin{theorem}
\label{salience payoff theorem}
    For any strategy $s_1^*$, if $\omega_{s_1^*} \in \Omega$ and $s_1^*$ is confound-defeating and has salience $\beta$, then
    \[ \liminf_{\delta \to 1} \und U_{1}(\delta) \geq \beta V(s_1^*) + (1 - \beta) V_0 (s_1^*). \]
\end{theorem}

Note that if $s_1^*$ is not behaviorally confounded then $\Omega_0(s_1^*)=\{\omega_{s_1^*},\omega_R\}$ for sufficiently small $\eta$. By upper hemi-continuity of $B(\cdot)$, this implies that $c_{\eta, \varsigma} = -\infty$ for sufficiently small $\eta$ and $\varsigma$, so $\beta = 1$. Theorem \ref{salience payoff theorem} therefore generalizes Theorem \ref{main payoff commitment theorem}.

Moreover, as $c_0(s_1^*)<1$, $\beta \to 1$ whenever $\mu_0(\omega_{s_1^*} | \Omega_0(s_1^*) \setminus \{\omega_R\}) \to 1$, in which case Theorem \ref{salience payoff theorem} recovers the conclusion of Theorem \ref{main payoff commitment theorem} even if $s_1^*$ is behaviorally confounded. But Theorem \ref{main payoff commitment theorem} delivers much more than continuity of the payoff lower bound at $\mu_0|_{\Omega_0(s_1^*)\setminus \{\omega_R\}}(s_1^*) = 1$---it gives an explicit lower bound that declines linearly with $\mu_0(\omega_{s_1^*} | \Omega_0(s_1^*) \setminus \{\omega_R\})$.\footnote{This feature contrasts with the approach of \cite{ely2008reputation}, who show that introducing a sufficiently high conditional probability of the Stackelberg type overturns \citeauthor{ElyValimaki03}'s (\citeyear{ElyValimaki03}) bad reputation result, but require an unbounded likelihood ratio between the Stackelberg type and the ``bad commitment type.''}

For example, consider the deterrence game from Section \ref{s:deterrence} with two commitment types: the pure Stackelberg type $(A,F)$ and the type $s_1$ that takes $A$ with probability $p$ for each signal. In this example, $c_0(A,F)$ is the probability such that the short-run player is indifferent between $C$ and $D$ when the long-run player plays $(A,F)$ with probability $c_0(A,F)$ and plays $s_1$ with probability $1-c_0$, which is given by $c_0(A,F) =\frac{pg+(1-p)l}{(2p-1)(1+g)}$. The salience of type $(A,F)$ is then $$\beta = \max\left\{\frac{\mu_0|_{\Omega_0(s_1^*)\setminus \{\omega_R\}}(A,F)- c_0(A,F)}{1 - c_0(A,F)}, 0 \right\},$$ and Theorem \ref{salience payoff theorem} implies that as $\delta \to 1$ the long-run player is assured a payoff of at least $\beta p +(1-\beta)(1-p)$. In particular, whenever $C$ is the unique best response to $(A,F)$, we have $c_0(A,F)<1$, so that as the prior weight on $(A,F)$ relative to $s_1$ increases, $\beta$ converges to $1$ and the long-run player is assured her pure Stackelberg payoff $p$.

\section{Discussion} \label{s:discussion}

This paper has studied reputation-formation when a player desires a reputation for conditional action. The main result is that if a strategy is \emph{confound-defeating} and either \emph{not behaviorally confounded} or \emph{salient}, a patient long-run player can secure the corresponding commitment payoff. A strategy is confound-defeating if and only if it is the unique solution to an optimal transport problem. In one-dimensional supermodular games, a strategy is confound-defeating if and only if it is monotone. In delegation games with supermodular agent utility or in signaling games with state-independent sender preferences and strictly submodular signaling costs, a patient agent or sender can secure her commitment payoff from any monotone strategy. Finally, we characterized communication strategies that are monotone with respect to some order and are thus implementable with a small ``lying cost.''

We mention some possible extensions of our results. First, the connection between unobserved deviations and optimal transport is not specific to the long-run/short-run model we study and could also be applied to repeated games with multiple long-run players, with or without incomplete information. Second, extending the model to allow multiple short-run players and to allow $u_0$ to depend on $a_2$ would encompass reputation-formation by a long-run mediator who coordinates play among multiple short-run players. This extension can potentially provide a reputational foundation for a general static mediation solution, as in \cite{myerson1982optimal}. Third, our results extend to the case with multiple rational types with different preferences, so long as they all have the same Stackelberg strategy and it is confound-defeating for all of them. A possible extension to the case with multiple rational types with different Stackelberg strategies that are confound-defeating only for some types would require additional analysis and qualifications. Fourth, cyclical monotonicity can be explored in multidimensional games, for example communication games with multidimensional states or actions. In particular, we are not sure if Proposition \ref{p:persuasion} has a useful multidimensional analogue. Fifth, in signaling games, we gave conditions under which the sender can secure her commitment payoff from any monotone strategy. An open question is when the Stackelberg signaling strategy is monotone. Finally, \cite{watson1993reputation} and \cite{battigalli1997reputation} show that the classic reputation results of \citeauthor{FudenbergLevine89} (\citeyear{FudenbergLevine89,FudenbergLevine92}) require only two rounds of iterated deletion of dominated strategies, rather than the full force of Nash equilibrium. Our results instead require three rounds of deletion: under our conditions, the long-run player secures the Stackelberg payoff by best responding to any short-run player best response to any long-run player strategy that is not undetectably dominated.

\appendix
\setcounter{section}{0}
\renewcommand{\thesection}{\Alph{section}}

\section{Appendix: Omitted Proofs} \label{s:proofs}

\subsection{Proof of Proposition \ref{p:CD}}

    If the second definition fails, there exist $(\alpha_0, \alpha_2) \in B_0 (s^*_1)$, $\varepsilon>0$, and a strategy $s_1'$ satisfying $||s_1' - s^*_1||>\varepsilon$ such that $\gamma(\alpha_0,s_1')$ solves $\text{OT}(\rho(\alpha_0),\phi(\alpha_0,s_1^*);\alpha_2)$. Since $B_0(s_1^*) \subset B_\eta(s_1^*)$ for all $\eta > 0$, this implies that the first definition fails.

    Conversely, if the first definition fails, there exists $\varepsilon > 0$ such that for all $\eta > 0$, there exist $s_1^\eta$ and $(\alpha_0^\eta, \alpha_2^\eta) \in B_\eta(s_1^*)$ where $||s_1^\eta - s_1^*|| > \varepsilon$, $||p(\alpha_0^\eta, s_1^\eta, \alpha_2^\eta) - p(\alpha_0^\eta, s_1^*, \alpha_2^\eta)|| < \eta$, and $s_1^\eta$ is not undetectably dominated: $u_1(\alpha_0^\eta, s_1^\eta, \alpha_2^\eta) \geq u_1(\alpha_0, \tilde s_1, \alpha_2^\eta)$ for all $\tilde s_1$ such that $p(\alpha_0^\eta, s_1^\eta, \alpha_2^\eta) = p(\alpha_0^\eta, \tilde s, \alpha_2^\eta)$. 
    Since $a_1$ is identified, whenever  $\phi(\alpha_0^\eta, s_1^\eta) = \phi(\alpha_0^\eta, \tilde s_1)$ we know $p(\alpha_0^\eta, s^\eta, \alpha_2^\eta)=p(\alpha_0^\eta, \tilde s, \alpha_2^\eta)$, and hence $s_1^\eta$ solves $\text{OT}(\rho(\alpha_0^\eta), \phi(\alpha_0^\eta, s_1^\eta); \alpha_2^\eta)$. 
   Now, since $B_\eta(s_1)\downarrow B_0(s_1)$, $\Delta(A_0) \times (\Delta(A_1))^{Y_0} \times \Delta(A_2)$ is compact, and $\text{OT}(\rho(\alpha_0),\phi(\alpha_0,s_1);\alpha_2)$ is jointly upper hemi-continuous in $(\alpha_0, s_1, \alpha_2)$, passing to the limit yields $s_1^0$ and $(\alpha_0, \alpha_2) \in B_0(s_1^*)$ such that $ ||s_1^0 - s_1^*|| \geq \varepsilon$ and $s_1^0$ solves $\text{OT}(\rho(\alpha_0), \phi(\alpha_0, s_1^*); \alpha_2)$. Thus, the second definition fails.

\subsection{Proof of Lemma \ref{l:gossner}} \label{gossner proof}

   For any two signal distributions $p$ and $q$, let $d(p || q) = \int \log\left(p/q \right) dp$ denote the relative entropy from $q$ to $p$. By Lemma 4 of \cite{Gossner11} and a standard application of the chain rule for relative entropy,
    \[\sum_t \mathbb{E}^{\mathbb{Q}}\left[ d(p(\sigma^*_0,s^*_1,\sigma^*_2|h^t) || p(\sigma^*_0, \bar \sigma^*_1,\sigma^*_2|h^t)) \right] \leq -\log \mu_0(\omega_{s_1^*}). \]
    Hence, by Markov's inequality,
    \[\mathbb{E}^{\mathbb{Q}} \left[ \# \left\{ t: d(p(\sigma^*_0,s^*_1,\sigma^*_2|h^t) || p(\sigma^*_0, \bar \sigma^*_1,\sigma^*_2|h^t)) > \frac{\eta^2}{2} \right\} \right] < -\frac{2 \log \mu_0(\omega_{s_1^*})}{\eta^2}. \]
    On the other hand, by Pinsker's inequality,
    \[  d\left( p(\sigma_0^*,s^*_1,\sigma_2^* | h^t) \bigg| \bigg| p(\sigma_0^*, \bar \sigma^*_1, \sigma_2^* | h^t) \right) \leq \frac{\eta^2}{2} \implies h^t \in H_\eta^t. \]
    This gives the desired bound.

\subsection{Proof of Lemma \ref{almost uniform convergence}} \label{uniform convergence proof}

We first show that the desired conclusion holds for each strategy profile $(\sigma_0^*, \sigma_1^*,\sigma_2^*)$ with $(\sigma_0^*,\sigma_2^*)\in B_1^H$ (and hence for any equilibrium), and then show that $\hat T$ can be fixed independent of the choice of $\delta$ and the equilibrium.

\begin{lemma}
    \label{l: pointwise convergence}
        For any strategy profile $(\sigma_0^*, \sigma_1^*,\sigma_2^*)$ where $(\sigma_0^*,\sigma_2^*)\in B_1^H$, and any $\zeta > 0$, there exists a set of infinite histories $G(\zeta) \subset H^\infty$ satisfying $\bb{Q}(G(\zeta)) > 1 - \zeta$ and a period $\hat T$ such that, for any $h \in G(\zeta)$ and any $t \geq \hat T$, we have $ \mu_t(\cdot | h) \in M_\zeta$. 
    \end{lemma}

\begin{proof}   
    
Since $\bb{Q}$ is absolutely continuous relative to $\overline{\bb{P}}$ and $\mu_t(\cdot | h)$ is a martingale relative to $\overline{\bb{P}}$, $\mu_t(\cdot | h)$ converges $\bb{Q}$-almost surely to some limit distribution $\mu_\infty(\cdot | h)$ (e.g., \cite{mailath2006repeated}, Lemma 15.4.2). 

We show that, for $\bb{Q}$-almost all histories $h\in H^\infty$, $\mu_\infty(\{\omega_R,\omega_{s_1^*}\} | h) = 1$. Suppose that $\mu_\infty(\Omega \setminus \{\omega_R\} | h) >0$, let $\omega_{s_1} \in \Omega \setminus \{\omega_R\}$ satisfy $\mu_\infty(\omega_{s_1} | h) >0$, and let $c>0$ and $T$ satisfy $\mu_t(\omega_{s_1} | h) >c$ for all $t>T$. Suppose also that the set of signals $y_1$ that realize infinitely often in $h$ is precisely $Y_1^*=\supp (\rho_1 (\cdot|s^*_1(Y_0),a_2))$ (which is independent of the choice of $a_2$ by Assumption \ref{a:signal}(2)), the set of signals that arise with positive probability when player 1 plays $s_1^*$. This supposition is without loss as $\rho_0(\cdot|a_0)$ has full support (by Assumption \ref{a:signal}(1)), so such histories occur $\bb{Q}$-almost surely. Next, for any $\Omega' \subseteq \Omega \setminus \{\omega_R\}$, let $p_{Y_1}(\sigma_0^*, \tilde{s}_1 | h^t, \Omega')$ denote the distribution of $y_1$ conditional on reaching history $h^t$ and the event $\omega \in \Omega'$; when $\Omega'$ is a singleton, $\Omega'=\{\hat{s}_1\}$, we write this as $p_{Y_1}(\sigma_0^*, \hat{s}_1 | h^t)$. Then, for any $y_1 \in Y_1^*$, we have
   \begin{align*}
       & \left|\mu_{t + 1}(\omega_{s_1} | \Omega \setminus \{\omega_R\}, h^t, y_1) - \mu_t(\omega_{s_1} | \Omega \setminus \{\omega_R\}, h^t)\right| \\
       =& \left| \frac{p_{Y_1}(\sigma_0^*, s_1 | h^t)[y_1]\mu_t(\omega_{s_1} | \Omega \setminus \{\omega_R\}, h^t)}{p_{Y_1}(\sigma_0^*, \tilde{s}_1 | h^t, \Omega \setminus \{\omega_R\})[y_1]} - \mu_t(\omega_{s_1} | \Omega \setminus \{\omega_R\}, h^t) \right|
        \\ =& \frac{\mu_t(\omega_{s_1} | \Omega \setminus \{\omega_R\}, h^t)}{p_{Y_1}(\sigma_0^*, \tilde{s}_1 | h^t, \Omega \setminus \{\omega_R\})[y_1]} \left| p_{Y_1}(\sigma_0^*, s_1| h^t)[y_1] - p_{Y_1}(\sigma_0^*, \tilde{s}_1 | h^t, \Omega \setminus \{\omega_R\})[y_1]  \right|
        \\ > & c \left|p_{Y_1}(\sigma_0^*, s_1 | h^t)[y_1] - p_{Y_1}(\sigma_0^*, \tilde{s}_1 | h^t, \Omega \setminus \{\omega_R\})[y_1]  \right|,
   \end{align*}
   where the first equality is by Bayes' rule, the second pulls out a common term, and the inequality is by the definition of $c$.
      Since $\mu_t(\cdot | h)$ converges, $y_1$ realizes infinitely often, and $c > 0$, this implies
    \[ \lims_{t \to \infty} \left| p_{Y_1}(\sigma_0^*, s_1 | h^t)[y_1] - p_{Y_1}(\sigma_0^*, \tilde{s}_1| h^t, \Omega \setminus \{\omega_R\})[y_1] \right|  = 0.\]
 At the same time, applying the argument in Lemma \ref{l:gossner} conditional on the event $\omega \neq \omega_R$ and the fact $s_1^*$ is not behaviorally confounded implies that, for $\bb{Q}$-almost all histories $h \in H^\infty$, 
    \[ \lims_{t \to \infty} \left| \left| p_{Y_1}(\sigma_0^*, \tilde{s}_1 | h^t, \Omega \setminus \{\omega_R\}) - p_{Y_1}(\sigma_0^*, s_1^* | h^t) \right| \right| = 0.\]
    In particular, since $p_{Y_1}(\sigma_0^*, s_1^* | h^t)[y_1]=0$ for all $y_1 \notin Y^*_1$, this implies that, for all $y_1 \notin Y^*_1$,
    \[\lim_{t\to \infty} p_{Y_1}(\sigma_0^*, \tilde{s}_1| h^t, \Omega \setminus \{\omega_R\})[y_1] =0.    \]
    Since we have already shown that $p_{Y_1}(\sigma_0^*, \tilde{s}_1| h^t, \Omega \setminus \{\omega_R\})[y_1]$ and $p_{Y_1}(\sigma_0^*, s_1 | h^t)[y_1]$ have the same limit for all $y_1 \in Y_1^*$, we have
         \[ \lims_{t \to \infty} \left\| p_{Y_1}(\sigma_0^*, s_1 | h^t) - p_{Y_1}(\sigma_0^*, \tilde{s}_1 | h^t, \Omega \setminus \{\omega_R\}) \right\|  = 0.\]
   Thus, by the triangle inequality,
   \[ \lims_{t \to \infty} \left| \left| p_{Y_1}(\sigma_0^*, s_1 | h^t) - p_{Y_1}(\sigma_0^*,s_1^* | h^t) \right| \right| = 0.   
   \]
    Finally, since $\omega_{s_1} \in \Omega$ and $s_1^*$ is not behaviorally confounded, this implies that $s_1=s_1^*$, and hence $\mu_\infty(\{\omega_R,\omega_{s_1^*}\} | h) = 1$.

To complete the proof, recall that Egorov's theorem shows that if a sequence of functions $f_n:H^\infty \to \bb{R}$ converges $\bb{Q}$-almost surely to $f$, then for all $\zeta >0$, there exists $G(\zeta) \subset H^\infty$ satisfying $\bb{Q}(G(\zeta)) \geq 1 - \zeta$ such that $f_n \to f$ uniformly on $G(\zeta)$. Thus, Lemma \ref{l: pointwise convergence} follows from Egorov's theorem applied to the sequence of conditional beliefs $\mu_t(\{\omega_{s_1^*}, \omega_R\}|h)$ and the definition of uniform convergence.
\end{proof}

We now show that $\hat T$ can be chosen as a function only of $\zeta$ and not of $\delta$ or the equilibrium strategies. To this end, let $\bb{Q}^{\sigma_0, \sigma_2}$ be the probability measure on $H^\infty$ induced by strategies $(\sigma_0, s_1^*, \sigma_2)$, and let $\mu_t^{\sigma_0, \sigma_2}(\omega_{s_1^*} | \Omega \setminus \{\omega_R\}, h^t)$ be the conditional belief that the long-run player is of type $\omega_{s_1^*}$ conditional on being irrational. This is well-defined $\bb{Q}^{\sigma_0, \sigma_2}$-almost surely because, conditional on the event $\Omega \setminus \{\omega_R\}$, the rational long-run player's strategy does not affect $\mu_t^{\sigma_0, \sigma_2}$ once $(\sigma_0, \sigma_2)$ are given. Next, let $L^{\sigma_0, \sigma_2} \subset H^\infty$ be the set of all histories $h$ where $\mu_t^{\sigma_0, \sigma_2}(\omega_{s_1^*} | \Omega \setminus \{\omega_R\}, h^t)  \to 1$ as $t \to \infty$. 
By Lemma \ref{l: pointwise convergence}, $\bb{Q}^{\sigma_0, \sigma_2} \left( L^{\sigma_0, \sigma_2} \right)  = 1$ for all $(\sigma_0, \sigma_2) \in B_1^H$. 

Following \citeauthor{mertens2015repeated} (\citeyear{mertens2015repeated}, p.\ 173), we identify $B_1^H$ with the space $(B_1^\infty)^{H^\infty}$ of maps from infinite histories $h$ to sequences of stage game strategies $(\alpha_0,\alpha_2)^\infty$. We show that $(B_1^\infty)^{H^\infty}$ is compact in the topology of uniform convergence. 
Endow $H^\infty=Y_1^\infty$ with the product topology, where each coordinate is endowed with the discrete topology; by Tychonoff's theorem, this makes $H^\infty$ compact. 
Endow $B_1^\infty$ with the product topology, which is generated by the metric 
\[ \tilde d((\alpha_0, \alpha_2)^\infty, (\alpha_0', \alpha_2')^\infty) = \sum_{t = 0}^\infty \frac{1}{2^t} ||(\alpha_0, \alpha_2)_t - (\alpha_0', \alpha_2')_t ||. \]
This is again compact by Tychonoff's theorem because $\Delta(A_0)\times\Delta(A_2)$ is compact. 
Note that each $(\sigma_0,\sigma_2):H^\infty \to B_1^\infty$ is continuous with these topologies, because if two sequences of histories agree for the first $T$ periods then the $\tilde d$-distance between their images under $(\sigma_0,\sigma_2)$ is at most $1/2^{T-1}$. Moreover, because the modulus of continuity is independent of both the choice of histories and the choice of $(\sigma_0,\sigma_2)$, the family of functions $(B_1^\infty)^{H^\infty}$ is equicontinuous. Next, endow $(B_1^\infty)^{H^\infty}$ with the topology of uniform convergence, which is generated by the $\sup$ norm 
\[ d((\sigma_0, \sigma_2), (\sigma_0', \sigma_2')) = \sup_{h \in H}  \tilde d((\sigma_0(h), \sigma_2(h)), (\sigma_0'(h), \sigma_2'(h))  . \]
Thus, $(B_1^\infty)^{H^\infty}$ is equicontinuous and closed.

Moreover, as $B_1^\infty$ is compact, for any $h\in H^\infty$ the closure of the set $\{(\sigma_0(h), \sigma_2(h))\}_{(\sigma_0, \sigma_2) \in (B_1^\infty)^{H^\infty}}$ is a compact subset of $B_1^\infty$. We thus satisfy the conditions of Theorem 47.1 in \cite{munkres2000} (generalized Arzela-Ascoli), implying that $(B_1^\infty)^{H^\infty}$ is compact in the topology of uniform convergence\footnote{Theorem 47.1 of \citeauthor{munkres2000} proves compactness of $(B_1^\infty)^{H^\infty}$ in the topology of compact convergence, which Theorem 46.7 of \citeauthor{munkres2000} shows coincides with the topology of uniform convergence when $H^\infty$ is compact.}. 

We can now prove that $\hat T(\zeta, \sigma_0^*, \sigma_2^*)$ can be chosen independent of the choice of $(\sigma^*_0, \sigma^*_2)$ (and hence also independent of $\delta$).
Suppose by contradiction that the statement is false. Then there exists  $\zeta > 0$ such that for each time $T \in \bb{N}$, there exist strategies $(\sigma_0^T, \sigma_2^T) \in B_1^H$ and a set of histories $E_{T}(\zeta)\subset H^\infty$ such that $\bb{Q}^{\sigma_0^T, \sigma_2^T}(E_{T}(\zeta)) > \zeta$ but $\mu_{T}(\cdot | h) \not\in M_\zeta$ for all $h \in E_{T}(\zeta)$.
Taking a subsequence if necessary and using compactness of $(B_1^\infty)^{H^\infty}$, we have $(\sigma_0^T, \sigma_2^T) \to (\sigma_0^\infty, \sigma_2^\infty) \in B_1^H$ in the topology of uniform convergence. 
Since $\mu_T$ depends only on the history up to time $T$, this implies that $\bb{Q}_{T'}^{\sigma_0^T, \sigma_2^T}(E_{T}(\zeta)) > \zeta$ for all $T' \geq T$, where $\bb{Q}_{T'}$ is the measure $\bb{Q}$ conditioned on the time-$T'$ $\sigma$-algebra. 
Thus, since $\bb{Q}_{T'}$ is continuous in strategies as a finite-dimensional measure,  passing $(\sigma_0^T, \sigma_2^T)$ to the limit (while fixing the time $T'$ and the set of histories $E_{T}(\zeta)$) gives $\bb{Q}_{T'}^{\sigma_0^\infty, \sigma_2^\infty}(E_{T}(\zeta)) > \zeta$ for all $T'$ sufficiently large; and then taking $T' \to \infty$ gives $\bb{Q}^{\sigma_0^\infty, \sigma_2^\infty}(E_{T}(\zeta)) > \zeta$. As this holds for all $T$, we have a sequence of events $\{E_{T}(\zeta)\}_{T \in \bb{N}}$ such that $\bb{Q}^{\sigma_0^\infty, \sigma_2^\infty}(E_{T}(\zeta)) > \zeta$ for all $T$. 
From here, we can conclude\footnote{This is a consequence of the Kochen-Stone Theorem; see, for example, Theorem 1.3 of \cite{Arthan2021}. We thank Eric Gao for pointing us to this result.}
\[ \bb{Q}^{\sigma_0^\infty, \sigma_2^\infty}\left( \limsup_{T \to \infty} E_{T}(\zeta) \right) \geq \limsup_{n \to \infty}  \frac{\left(\sum_{k = 1}^n \bb{Q}^{\sigma_0^\infty, \sigma_2^\infty}(E_{k}(\zeta))\right)^2}{\sum_{1 \leq j, k \leq T'} \bb{Q}^{\sigma_0^\infty, \sigma_2^\infty}(E_j(\zeta) \cap E_k(\zeta))} > \frac{n^2 \zeta^2}{n^2} = \zeta^2. \]
Thus, for any history $h \in \limsup_{T \to \infty} E_{T}(\zeta) = E_\infty(\zeta)$, there is a sequence of times $\{T_n\}$ such that $\mu_{T_n}(\cdot | h) \notin M_\zeta$ for all $n$. Since $\zeta^2 < \zeta$, this implies that $\mu_{T_n}(\cdot | h) \notin M_{\zeta^2}$. Thus, for any $h \in E_\infty(\zeta)$,  $\mu_t(\cdot | h) \notin M_{\zeta^2}$ for infinitely many $T$; but $\bb{Q}^{\sigma_0^\infty, \sigma_2^\infty}(E_\infty(\zeta)) > \zeta^2$. But this contradicts Lemma \ref{l: pointwise convergence} for the strategies $(\sigma_0^\infty, \sigma_2^\infty)\in B_1^H$, completing the proof.

\subsection{Proof of Lemma \ref{almost close}} \label{almost close proof}
Fix any $h^t \in H_\eta^t$ and $\mu_t (\cdot|h^t) \in M_0$.
Then  
\[ \left| \left| p(\sigma_0^*, \bar \sigma_1^*, 
\sigma_2^* | h^t) - p(\sigma_0^*, s_1^*, \sigma_2^* | h^t)\right| \right|  = |\mu_t(\omega_R | h^t)| \left| \left| p(\sigma_0^*, \sigma_1^*, 
\sigma_2^* | h^t) - p(\sigma_0^*, s_1^*, \sigma_2^* | h^t)\right| \right| < \eta. \]

Since $p(\sigma_0^*, s_1^*, \sigma_2^* | h^t)$ is continuous in $\mu_t (\cdot|h^t)$, there exists a strictly positive function $\zeta(\eta)$ satisfying $\lim_{\eta \to 0}\zeta(\eta)=0$ such that if $h^t \in H_\eta^t$ and $\mu_t (\cdot|h^t) \in M_{\zeta(\eta)}$ then 
\begin{align*}
    |\mu_t(\omega_R | h^t)| \left| \left| p(\sigma_0^*, \sigma_1^*, \sigma_2^* | h^t) - p(\sigma_0^*, s_1^*, \sigma_2^* | h^t)\right| \right| < 2\eta
\end{align*}
and $(\sigma^*_0(h^t),\sigma^*_2(h^t)) \in B_{2\eta}(s^*_1)$.

Now fix $c: (0, 1] \to (0, 1]$ satisfying $c(\eta) \to 0, \frac{\eta}{c(\eta)} \to 0$ as $\eta \to 0$. 
If $\mu_t(\omega_R | h^t) \geq c(\eta)$ then
\[ \left| \left| p(\sigma_0^*, \sigma_1^*, \sigma_2^* | h^t) -  p(\sigma_0^*, s_1^*, \sigma_2^* | h^t)\right| \right| < \frac{2\eta}{c(\eta)}, \]
and in addition $h^t \in \mathrm{supp}(\mathbb{P})$. Hence, as $\eta \to 0$, Lemma \ref{confound defeating payoff} implies that $||\sigma_1^*(h^t) - s_1^*|| \to 0$, and so $(\sigma^*_0(h^t),\sigma^*_2(h^t)) \in \hat{B}_{\xi_1(\eta)}(s^*_1)$ for some strictly positive function $\xi_1(\eta)$ satisfying $\xi_1(\eta) \to 0$. If instead $\mu_t(\omega_R | h^t) < c(\eta)$ then $||\bar \sigma^*_1(h^t)- s_1^*|| \leq 1 - \zeta(\eta) - c(\eta)$, and hence $(\sigma^*_0(h^t),\sigma^*_2(h^t)) \in \hat{B}_{\zeta(\eta)+c(\eta)}(s^*_1)$. Taking $\xi(\eta) = \max\{\xi_1(\eta), \zeta(\eta) + c(\eta)\}$ completes the proof.

\subsection{Proof of Proposition \ref{p:cyclical}}

    The forward direction. Let $\gamma$ be feasible in $\text{OT}(\rho,\phi)$ with $\supp(\gamma)$ not strictly $u$-cyclically monotone. Let $\{(x_i, y_i)_{i = 1}^N\} \subset \supp(\gamma)$ be a collection of pairs witnessing a violation and set $\varepsilon=\min_{i} \gamma((x_i,y_i))$. Define $\gamma' \in \Delta(X \times Y)$ by 
    \begin{align*}
        \gamma'(x,y)=
           \begin{cases}
        \gamma(x,y)-\varepsilon & \text{if } (x,y)\in \left\{(x_i, y_i)_{i=1}^N\right\} \setminus \left\{(x_i, y_{i+1})_{i=1}^N\right\}, \\
        \gamma(x,y)+\varepsilon & \text{if } (x,y)\in \left\{(x_i, y_{i+1})_{i=1}^N\right\} \setminus \left\{(x_i, y_i)_{i=1}^N\right\}, \\
        \gamma(x,y) & \text{otherwise}.
    \end{cases}
    \end{align*}
    Then $\gamma' \neq \gamma$ is feasible in $\text{OT}(\rho,\phi)$ (since $\gamma$ and $\gamma'$ transport the same mass into and out of each point) and $\int u(x,y) d\gamma \leq \int u(x,y) d\gamma'$ (since $\{(x_i, y_i)_{i = 1}^N\}$ witnesses a violation of strict $u$-cyclical monotonicity), so $\gamma$ does not uniquely solve $\text{OT}(\rho,\phi)$.

    Conversely, if $\gamma$ is feasible in $\text{OT}(\rho,\phi)$ and $\supp(\gamma)$ is strictly $u$-cyclically monotone, consider any feasible $\gamma'\neq \gamma$ in $\text{OT}(\rho,\phi)$. Since $\gamma$ and $\gamma'$ are both feasible and $\gamma \neq \gamma'$, there exists $\{(x_i, y_i)_{i = 1}^N\} \subset \supp(\gamma)$ such that $\{(x_i, y_{i+1})_{i = 1}^N\} \subset \supp(\gamma')$. (To see this, let $(x_1,y_1)$ be any pair such that $\gamma(x_1,y_1)>\gamma'(x_1,y_1)$. Since $\gamma$ and $\gamma'$ transport the same mass into $y_1$, there exists $x_2$ such that $\gamma(x_2,y_1)<\gamma'(x_2,y_1)$. But now, since $\gamma$ and $\gamma'$ transport the same mass out of $x_2$, there exists $y_2$ such that $\gamma(x_2,y_2)>\gamma'(x_2,y_2)$. Continuing in this manner and using finiteness of $X \times Y$ yields a cycle.) Let $\varepsilon=\min_i \gamma'(x_i,y_{i+1})$, and let
    \begin{align*}
        \gamma''(x,y)=
           \begin{cases}
        \gamma'(x,y)-\varepsilon & \text{if } (x,y)\in \left\{(x_i, y_{i+1})_{i=1}^N\right\} \setminus \left\{(x_i, y_{i})_{i=1}^N\right\}, \\
        \gamma'(x,y)+\varepsilon & \text{if } (x,y)\in \left\{(x_i, y_{i})_{i=1}^N\right\} \setminus \left\{(x_i, y_{i+1})_{i=1}^N\right\}, \\
        \gamma'(x,y) & \text{otherwise}.
    \end{cases}
    \end{align*}
 Then $\gamma''$ is feasible in $\text{OT}(\rho,\phi)$ and $\int u(x,y) d\gamma'' > \int u(x,y) d\gamma'$ (since $\left\{(x_i, y_i)_{i=1}^N\right\}$ is contained in the strictly $u$-cyclically monotone set $\supp(\gamma)$), so $\gamma'$ does not solve $\text{OT}(\rho,\phi)$. Thus, since no $\gamma' \neq \gamma$ solves $\text{OT}(\rho,\phi)$, and $\text{OT}(\rho,\phi)$ has a solution as a continuous maximization problem over a compact set, $\gamma$ must uniquely solve $\text{OT}(\rho,\phi)$.

\subsection{Proof of Proposition \ref{p:converse}}

The result is obvious if $\bar v_1^{\mathrm{CM}} = \bar u_1$, so suppose $\bar v_1^{\mathrm{CM}}<\bar{u}_1$, and fix $\varepsilon<2(\bar{u}_1-\bar v_1^{\mathrm{CM}})$, a prior $\mu_0$ with $\mu_0(\omega_R) > 0$, and an equilibrium $(\sigma_0^*,\sigma_1^*,\sigma_2^*)$. We start with an additional lemma.

\begin{lemma}
    \label{uniform upper semi-continuity}
   For all $\xi > 0$, there exists $\varsigma > 0$ such that, for any $u_1$-cyclically monotone strategy $s_1$, any strategy $s_1'$ satisfying $||s_1 - s_1'|| < \varsigma$, and any $(\alpha_0,\alpha_2)\in B(s'_1)$, we have $u_1(\alpha_0, s_1, \alpha_2) < \bar v_1^{\mathrm{CM}} + \xi$.
\end{lemma}
\begin{proof}
Suppose for contradiction that there exists $\xi>0$ and a sequence of strategies $\tilde s_1^n$, each within $1/n$ of a $u_1$-cyclically monotone strategy $s_1^{n}$, and $(\alpha_0^n,\alpha_2^n)\in B(\tilde s_1^n)$ such that $u_1(\alpha_0^n, s_1^n, \alpha_2^n) \geq \bar v_1^{\mathrm{CM}} + \xi$. Taking a subsequence if necessary and noting that the set of $u_1$-cyclically monotone strategies is closed  (as they are characterized by their support) and hence compact, $s_1^{n}$ converges to a $u_1$-cyclically monotone strategy $s_1$. Moreover, $\tilde s_1^n$ also converges to $s_1$, by the triangle inequality. But this yields a contradiction, as
   \[ \bar v_1^{\mathrm{CM}} + \xi \leq \limsup_{n \to \infty} u_1(\alpha_0^n, s_1^{n}, \alpha_2^n) \leq \sup_{(\alpha_0,\alpha_2)\in B(s_1)} u_1(\alpha_0, s_1, \alpha_2) \leq \bar v_1^{\mathrm{CM}}. \]
   The first inequality is by definition of $\tilde s_1^n$, the second inequality follows because $B(\cdot)$ is upper hemi-continuous and $u_1$ is continuous, and the third inequality follows because $s_1$ is $u_1$-cyclically monotone.
\end{proof}

Now, note that at any history $h^t$ where $\mu_t(\omega_R | h^t) > 1 - \varsigma$, we have $||\bar \sigma_1^*(h^t)-\sigma_1^*(h^t)||<\varsigma$. Thus, by Lemmas \ref{must be monotone} and \ref{uniform upper semi-continuity}, there exists $\varsigma>0$ such that at any history $h^t$ where $\mu_t(\omega_R | h^t) > 1 - \varsigma$, we have $u_1(\sigma_0^*,\sigma_1^*,\sigma_2^*|h^t) < \bar v_1^{\mathrm{CM}}+\varepsilon/2$. Since $\mu_t(\omega_R|h^t,\omega_R)$ is a $\bb{P}$-submartingale,
\begin{align*}
    (1-\bb{P}(\mu_t(\omega_R | h^t,\omega_R)>1-\varsigma))(1-\varsigma)+\bb{P}(\mu_t(\omega_R | h^t,\omega_R)>1-\varsigma)(1) &\geq \mu_0(\omega_R) \quad \Longleftrightarrow \\
    \bb{P}(\mu_t(\omega_R | h^t,\omega_R)>1-\varsigma) &\geq 1-\frac{1-\mu_0(\omega_R)}{\varsigma}. 
\end{align*}
Therefore, the (rational) long-run player's expected payoff in each period $t$ is at most
\[\left(1-\frac{1-\mu_0(\omega_R)}{\varsigma}\right)\left(\bar v_1^{\mathrm{CM}}+\frac{\varepsilon}{2}\right)+\frac{1-\mu_0(\omega_R)}{\varsigma}\bar{u}_1.\]
This completes this proof, as this payoff is less than $\bar v_1^{\mathrm{CM}}+\varepsilon$ whenever \[\mu_0(\omega_R)>1-\frac{\varepsilon \varsigma}{2(\bar{u}_1-\bar v_1^{\mathrm{CM}})-\varepsilon}.\]

\subsection{Proof of Proposition \ref{p:chain}}

Suppose that $s_1:\Theta \to \Delta(R)$ is monotone with respect to $(\succsim_\Theta,\succsim_R)$. 

First, $G(s_1)$ cannot contain a cycle $(\theta_1,r_1),(\theta_2,r_1),\ldots,(\theta_K,r_K),(\theta_1,r_K)$. To see this, suppose otherwise, and let $\theta_1 \prec \ldots \prec \theta_K$, without loss. Since $r_k \in \supp(s_1(\theta_k)) \cap \supp(s_1(\theta_{k+1}))$ for $k\in \{1,\ldots,K-1\}$ and $r_K \in \supp(s_1(\theta_K))$, monotonicity requires $r_1 \prec \ldots \prec r_K$. But this gives a contradiction, since $r_K \in \supp (s_1(\theta_1))$ and $r_1 \in \supp(s_1(\theta_2))$.

Next, $G(s_1)$ cannot contain a forbidden triple. We have already explained why it cannot contain a Type (1) forbidden triple. The argument for why it cannot contain a Type (2) forbidden triple is identical, with the roles of states and actions interchanged.

Conversely, suppose that $G(s_1)$ is acyclic and does not contain a forbidden triple. It suffices to consider the case where $G(s_1)$ is connected, as otherwise the orders on the states and actions in each connected component of $G(s_1)$ can be appended to one another. So suppose that $G(s_1)$ is connected, and let $(\theta_1,r_1),(\theta_2,r_1),\ldots,(\theta_K,r_K)$ be any maximum path in $G(s_1)$ (i.e., any path of maximum length), supposing that such a path ends with an action. (The argument for the case where all maximum paths both start and end at actions or at states is almost identical.) Define the orders $\prec_\Theta$ on $\{\theta_1,\ldots,\theta_K\}$ and $\prec_R$ on $\{r_1,\ldots,r_K\}$ by $\theta_1 \prec_\Theta \ldots \prec_\Theta \theta_K$ and $r_1 \prec_R \ldots \prec_R r_K$.

We claim that for any $\theta \in \Theta \setminus \{\theta_1,\ldots,\theta_K\}$, there exists $k \in \{1,\ldots, K-1\}$ such that $\supp(s_1(\theta))=\{r_k\}$. Note such a state $\theta$ is linked to at most one $r_k \in \{r_1,\ldots, r_{K-1}\}$, as if $\theta$ is linked to distinct $r_k,r'_k$ then appending $\theta$ to both ends of the path from $r_k$ to $r'_k$ forms a cycle. In addition, $\theta$ cannot be linked to $r_K$, as then it could be appended to the maximum path. Finally, $\theta$ cannot be linked to both some $r_k \in \{r_1,\ldots, r_{K-1}\}$ and some $r \in R \setminus \{r_1,\ldots, r_{K-1}\}$. For, if $k=1$ then replacing $(\theta_1,r_1)$ with $(\theta,r),(\theta,r_1)$ at the beginning of the maximum path would lengthen it; and if $k\geq 2$ then the set of states $\{\theta_k,\theta,\theta_{k+1}\}$ together with the set of actions $\{r,r_{k-1},r_k,r_{k+1}\}$ would be a forbidden triple, as $\{r_{k-1},r_k\}\in \supp(s_1(\theta_k))$, $\{r,r_k\}\in \supp(s_1(\theta))$, and $\{r_{k},r_{k+1}\}\in \supp(s_1(\theta_{k+1}))$ (see Figure \ref{f:triple}).

\begin{figure}[h]
    \centering 
\begin{minipage}{0.45\textwidth}

\begin{tikzpicture}[thick,
  every node/.style={draw,circle},
  fsnode/.style={fill=myblue},
  ssnode/.style={fill=mygreen},
  every fit/.style={ellipse,draw,inner sep=-2pt,text width=2cm},
  ->,shorten >= 3pt,shorten <= 3pt
]

\begin{scope}[start chain=going below,node distance=7mm]
    \node[fsnode, on chain] (f1) [label = left: $\theta_1$] {};
    \node[fsnode, on chain] (f2) [label = left: $\theta_2$] {};
        \node[fsnode, on chain] (f4) [label = left: $\theta$] {};
    \node[fsnode,on chain] (f3) [label=left: $\theta_{3}$] {};
\end{scope}

\begin{scope}[xshift=4cm,yshift=-0.5cm,start chain=going below,node distance=7mm]
\foreach \i in {1, ..., 3}
  \node[ssnode,on chain] (s\i) [label=right: $r_{\i}$] {};

  \node[ssnode, on chain] (s4) [label = right:$r$] {};
\end{scope}

\node [myblue,fit=(f1) (f3),label=above:$\Theta$] {};
\node [mygreen,fit=(s1) (s4),label=above:$R$] {};

\draw (f1) -- (s1);
\draw (f2) -- (s1);
\draw (f2) -- (s2);
\draw (f3) -- (s2);
\draw (f3) -- (s3);
\draw (f4) -- (s2);

\end{tikzpicture}
\end{minipage}
\hfill
\begin{minipage}{0.45\textwidth}
    
\begin{tikzpicture}[thick,
  every node/.style={draw,circle},
  fsnode/.style={fill=myblue},
  ssnode/.style={fill=mygreen},
  every fit/.style={ellipse,draw,inner sep=-2pt,text width=2cm},
  ->,shorten >= 3pt,shorten <= 3pt
]

\begin{scope}[start chain=going below,node distance=7mm]
    \node[fsnode, on chain] (f1) [label = left: $\theta_1$] {};
    \node[fsnode, on chain] (f2) [label = left: $\theta_2$] {};
        \node[fsnode, on chain] (f4) [label = left: $\theta$] {};

    \node[fsnode,on chain] (f3) [label=left: $\theta_{3}$] {};
\end{scope}

\begin{scope}[xshift=4cm,yshift=-0.5cm,start chain=going below,node distance=7mm]
\foreach \i in {1, ..., 3}
  \node[ssnode,on chain] (s\i) [label=right: $r_{\i}$] {};

  \node[ssnode, on chain] (s4) [label = right:$r$] {};
\end{scope}

\node [myblue,fit=(f1) (f3),label=above:$\Theta$] {};

\node [mygreen,fit=(s1) (s4),label=above:$R$] {};

\draw (f1) -- (s1);
\draw[red] (f2) -- (s1);
\draw[red] (f2) -- (s2);
\draw[red] (f3) -- (s2);
\draw[red] (f3) -- (s3);
\draw[red] (f4) -- (s2);
\draw[red] (f4) -- (s4);

\end{tikzpicture}

\end{minipage}
\caption{Each State $\theta \notin \{\theta_1,\ldots,\theta_K\}$ Has Only One Neighbor}
\caption*{\emph{Notes.}
If $\theta \notin \{\theta_1,\ldots,\theta_K\}$ is linked to $r_k \in \{r_1,\ldots,r_K\}$ and $r \notin \{r_1,\ldots,r_K\}$, then $\{\theta_k, \theta, \theta_{k+1}\}$ together with $\{r,r_{k-1}, r_k,r_{k+1}\}$ is a forbidden triple.}
\label{f:triple}
\end{figure}

Given this claim, we can extend $\prec_\Theta$ to $\Theta$ by ordering each $\theta \in \Theta \setminus \{\theta_1,\ldots,\theta_K\}$ such that $\supp(s_1(\theta))=\{r_k\}$ in between $\theta_k$ and $\theta_{k+1}$ (and ordering multiple such states arbitrarily between $\theta_k$ and $\theta_{k+1}$).

Similarly, for any $r \in R \setminus \{r_1,\ldots,r_K\}$, there exists $k \in \{2,\ldots, K\}$ such that $r_k$ is linked only to $\theta_k$ in $G(s_1)$. Extend $\prec_R$ to $R$ by ordering each such $r$ in between $r_{k-1}$ and $r_k$.

Note that for any $k \geq 2$ and any $r \in \supp(s_1(\theta_k))$, we have $r_{k-1} \precsim_R r \precsim_R r_k$. This follows because if $r \notin  \{r_1,\ldots, r_K\}$ then $r_{k-1} \precsim_R r \precsim_R r_k$ by construction, and if $r=r_{\tilde k}$ for some $\tilde k \notin \{k-1,k\}$, then $G(s_1)$ contains a cycle starting with $(\theta_k,r_{\tilde k})$ and then following the maximum path back to $\theta_k$

Finally, we claim that $s_1$ is monotone with respect to $(\succsim_\Theta,\succsim_R)$. To see this, fix any $\theta \succ_\Theta \theta'$. Let $\tilde k = \max \{k: \theta \succsim_\Theta \theta_k \}$. If $\theta \succ_\Theta \theta_{\tilde k} \succsim_\Theta \theta'$, then $\supp(s_1(\theta)) = \{r_{\tilde k}\}$ and $r_{\tilde k} \succsim_R r$ for all $r \in \supp(s_1(\theta'))$. If $\theta = \theta_{\tilde k} \succ_\Theta \theta'$, then $r_{\tilde k-1}$ is the lowest action in $\supp(s_1(\theta))$, and no action in $\supp(s_1(\theta'))$ is above $r_{\tilde k-1}$. Lastly, if $\theta \succ_\Theta \theta' \succsim_\Theta \theta_{\tilde k}$, then $\supp(s_1(\theta))=\supp(s_1(\theta'))=\{r_{\tilde k}\}$. Thus, in all cases, monotonicity holds.

\subsection{Proof of Theorem \ref{salience payoff theorem}}

The proof of Theorem \ref{salience payoff theorem} follows from the proof of Theorem \ref{main payoff commitment theorem} and the fact that $(\alpha_0,\alpha_2)\in B_\eta (s_1^*)$ (and hence $u_1(\sigma_0^*,s_1^*,\sigma_2^*)\geq \inf_{ (\alpha_0, \alpha_2) \in B_\eta(s^*_1) } u_1(\alpha_0, s^*_1, \alpha_2)$) for all $h^t \in H_\eta^t$, once we replace Lemma \ref{almost close} with the following lemma. 

\begin{lemma}\label{l:salience}
For any $\eta$ sufficiently small, any $t > \bar T(\eta)$ (chosen as in Theorem \ref{main payoff commitment theorem}), and any Nash equilibrium $(\sigma_0^*, \sigma_1^*, \sigma_2^*)$, we have 
    \[ \liminf_{\xi \to 0} \bb{Q}(h \in H^\infty : (\sigma_0^*(h^t), \sigma_2^*(h^t)) \in \hat B_{\xi}(s_1^*)) \geq \beta . \]
\end{lemma}

The proof of Lemma \ref{l:salience} in turn relies on the following technical lemma, which will be used to show that if $B(\mu) \subset B(s_1^*)$ and $d(\tilde \mu, \text{conv}(\{\mu,s_1^*\})) <\vartheta$ (where $d(\cdot,\cdot)$ denotes distance from a point to a set, and conv$(\cdot)$ denotes convex hull), then $B(\tilde \mu) \subset B(s_1^*)$, where $\vartheta$ can be chosen uniformly over a certain set of beliefs $\mu$.

\begin{lemma}
\label{exact best replies}
For any $s_1 \in \Delta (A_1)^{Y_0}$ and $\varsigma >0$, let 
\[ C_\varsigma(s_{1}) = \{s_{1}': B(s_{1}'') \subset B(s_{1}) \text{  for all   } s_{1}'' \text{ s.t. } \left\| s_{1}'-s_{1}'' \right\| \leq \varsigma \} .\]
Then, there exists $\vartheta(\varsigma, s_{1})>0$, vanishing as $\varsigma \to 0$, such that for all $\tilde s_{1}, s_{1}'$ such that $s_{1}' \in C_\varsigma(s_{1})$ and $d(\tilde s_{1}, \text{conv}(\{s_{1}, s_{1}'\})) \leq \vartheta(\varsigma, s_{1})$, we have $B(\tilde s_{1}) \subset B(s_{1})$. 
\end{lemma}
\begin{proof}
We first show that $B(\tilde s_{1}) \subset B(s_{1})$ if $\vartheta(\varsigma, s_{1})=0$, so that $\tilde s_{1} \in \text{conv}(\{s_{1}, s_{1}'\})$. To see this, note that since $B(s_{1}') \subset B(s_{1})$, the set of player $0$ best responses at any $\tilde s_{1} \in \text{conv}(\{s_{1}, s_{1}'\})$ (other than $s_{1}'$) is the same as that at $s_{1}'$, by the sure-thing principle. This then implies the same conclusion for player 2, so $B(\tilde s_{1})=B(s_{1}') \subset B(s_{1})$.

Next, we show there exists some $\vartheta(\varsigma, s_{1}) > 0$, which can be chosen independently of $s_{1}' \in C_{\varsigma}(s_{1})$, such that if $d(\tilde s_{1}, \text{conv}(\{s_{1}, s_{1}'\})) \leq \vartheta(\varsigma, s_{1})$ then $B(\tilde s_{1}) \subset B(s_{1})$. 
For any $\varsigma$, note that if $s_{1}'' \in \overline{C_\varsigma(s_{1})}$ (the closure of $C_\varsigma(s_{1})$), then $B(s_{1}'') \subset B(s_{1})$. 
From here, suppose no such $\vartheta > 0$ has the desired property for all $s_{1}' \in C_\varsigma(s_{1})$. Then there exists a sequence $(s_{1}^n, \tilde s_{1}^n)$ such that $s_{1}^n \in  C_\varsigma(s_{1})$, $d(\tilde s_{1}^n, \text{conv}(s_{1}^n, s_{1})) < \frac1n$, and $B(\tilde s_{1}^n) \setminus B(s_{1}) \neq \emptyset$ for large enough $n$. Taking a subsequence, this implies there exists $\left(s_{1}^{n_k}, \tilde s_{1}^{n_k}\right) \to (s_{1}', \tilde s_{1})$ 
such that $s_{1}' \in \overline{C_\varsigma(s_{1})}$ (since this set is closed) and $d(\tilde s_{1}, \text{conv}(s_{1}', s_{1})) = 0$, but $B(\tilde s_{1}) \setminus B(s_{1}) \neq \emptyset$. But this contradicts the fact, established above, that $B(\tilde s_{1}) \subset B(s_{1})$ when $\tilde s_{1} \in \text{conv}(\{s_{1}, s_{1}'\})$, completing the proof.
\end{proof}

\begin{proof}[Proof of Lemma \ref{l:salience}]
We show there exists strictly positive functions $\zeta(\eta)$ and $\xi(\eta)$, vanishing as $\eta \to 0$, and $\bar T(\eta)$ such that, for all $t > \bar T(\eta)$, 
\begin{align*}
\bb{Q}(h \in H^\infty : (\sigma_0^*(h^t), \sigma_2^*(h^t)) \in \hat B_{\xi(\eta)}(s_1^*))) &\geq (1-\zeta(\eta))\beta_{\zeta,\eta}, \; \text{ where} \\
\beta_{\varsigma, \eta}  &= \frac{(1-\eta)\mu_0(\omega_{s_1^*} | \Omega_0(s_1^*) \setminus \{\omega_R\})(s_1^*) - c_{\eta, \varsigma}}{1-\eta - c_{\eta, \varsigma}}.
\end{align*}
Lemma \ref{l:gossner} and an appropriate modification of Lemma \ref{almost uniform convergence} (with $\Omega_\eta (s_1^*)$ in place of $\{\omega_{s_1^*}\}$) imply that, on a set of histories $G(\zeta(\eta))$ satisfying $\bb{Q}(G(\zeta(\eta))) > 1 - \zeta(\eta)$, both $h^t \in H_\eta^t$ and $\mu_t(\Omega_\eta (s_1^*) \setminus \{\omega_R\} | h^t) > 1 - \eta$ for all $t > \bar{T}(\eta)$, independent of the choice of the equilibrium strategy and discount factor. Suppose these two conditions are satisfied and $t>\bar{T}(\eta)$. We consider three possible cases, and show for sufficiently small $\eta$, that in the first two cases $(\sigma_0^*(h^t), \sigma_2^*(h^t)) \in \hat B_{\xi(\eta)}(s_1^*)$ and the third arises with probability at most $1-\beta_{\varsigma,\eta}$. This then implies, in total, $\bb{Q}(h \in H^\infty : (\sigma_0^*(h^t), \sigma_2^*(h^t)) \in \hat B_{\xi(\eta)}(s_1^*)))$ is no less than $(1-\zeta(\eta))\beta_{\zeta,\eta}$, completing the proof.  

First, suppose that $\mu_t(\{\omega_R, \omega_{s_1^*}\}|h^t) > 1 - \zeta(\eta)$. Then, for $\zeta(\eta)$ and $\xi(\eta)$ chosen as in Lemma \ref{almost close}, we have that $(\sigma_0^*(h^t), \sigma_2^*(h^t)) \in \hat B_{\xi(\eta)}(s_1^*)$. 

Second, suppose that  $\mu_t(\{\omega_R, \omega_{s_1^*}\} | h^t) \leq 1 - \zeta(\eta)$ but
$\mu_t(\cdot | h^t, \Omega \setminus \{\omega_R\}) \in C_{\varsigma}(s_1^*)$, for some $\varsigma>0$ fixed independent of $\eta$. Since $h^t \in H_\eta^t$ and $\mu_t(\Omega_\eta (s_1^*)| h^t) > \mu_t(\Omega_\eta(s_1^*) \setminus \{\omega_R\}) > 1 - \eta$, an argument identical to the proof of Lemma \ref{almost close} implies that the minimum of $\mu_t(\omega_R | h^t)$ and $||\sigma_1^*(h^t) - s_1^*||$ is bounded above by a function that vanishes as $\eta \to 0$.
Thus, as $\eta$ vanishes, $d(\sigma_1^*(h^t), \text{conv}(\{\mu_t(\cdot | h^t, \Omega_\eta(s_1^*) \setminus \{\omega_R\}), s_1^*\})) \leq \vartheta(\varsigma, s_1^*)$ where $\vartheta(\cdot,\cdot)$ is defined in Lemma \ref{exact best replies}. Since $\mu_t(\cdot | h^t, \Omega \setminus \{\omega_R\}) \in C_{\varsigma}(s_1^*)$, Lemma \ref{exact best replies} then implies $(\sigma_0^*(h^t), \sigma_2^*(h^t)) \in B(s_1^*) \subset \hat B_{\xi(\eta)}(s_1^*)$.

Third, suppose that $\mu_t(\{\omega_R, \omega_{s_1^*}\}|h^t) \leq 1 - \zeta(\eta)$ and $\mu_t(\cdot | h^t, \Omega \setminus \{\omega_R\}) \not\in C_\varsigma(s_1^*)$. The former condition implies that $\mu_t(\cdot|h^t)$ satisfies Condition (1) of Definition \ref{confounding weights}, while the latter condition implies that it also satisfies Condition (2). Moreover, since $\mu_t(\Omega_\eta (s_1^*) | h^t) \geq 1 - \eta$ (as $t>\bar T (\eta)$), it also satisfies Condition (3). Thus, by the definition of $(\eta, \varsigma)$-confounding weights, $\mu_t(\omega_{s_1^*} | h^t,\Omega \setminus \{\omega_R\}) \leq c_{\varsigma, \eta}$. Now, since $\omega_{s_1^*} \in \Omega_\eta(s_1^*) \setminus \{\omega_R\}  \subset \Omega \setminus \{\omega_R\}$, we have 
\begin{align*}
    \mu_t(\omega_{s_1^*} | h^t, \Omega \setminus \{\omega_R\}) = \mu_t(\omega_{s_1^*} | h^t, \Omega_\eta(s_1^*) \setminus \{\omega_R\} ) \mu_t(\Omega_\eta(s_1^*) \setminus \{\omega_R\}  | h^t).
\end{align*}

Since $\mu_t(\omega_{s_1^*} | h^t, \Omega \setminus \{\omega_R\}) \leq c_{\varsigma, \eta}$ and $\mu_t(\Omega_\eta (s_1^*)  \setminus \{\omega_R\}  | h^t) \geq 1 - \eta$, we have
\[ \mu_t(\omega_{s_1^*} | h^t, \Omega_\eta(s_1^*)  \setminus \{\omega_R\} ) \leq \frac{c_{\varsigma, \eta}}{1 - \eta}.\]
Hence, the probability that $h$ lies in this third case is at most
\[
q_{\eta,\varsigma} := \bb{Q}\left(h \in H^\infty : \mu_t(\omega_{s_1^*}| h^t, \Omega_\eta(s_1^*)  \setminus \{\omega_R\}) \leq \frac{c_{\eta,\varsigma}}{1 - \eta} \right).
\]
Thus, because $\mu_t(\omega_{s_1^*} |h^t, \Omega_\eta(s_1^*) \setminus \{\omega_R\})$ is a $\bb{Q}$-submartingale, we have
\begin{align*}
    q_{\eta,\varsigma}\left(\frac{c_{\eta,\varsigma}}{1 - \eta}\right)+(1 - q_{\eta,\varsigma})(1) & \geq \mu_0(\omega_{s_1^*} | \Omega_\eta(s_1^*) \setminus \{\omega_R\}) \qquad \qquad \Longleftrightarrow \\
    q_{\eta,\varsigma} &\leq \min\left\{\frac{1- \mu_0(\omega_{s_1^*} | \Omega_\eta(s_1^*) \setminus \{\omega_R\})}{1-\frac{c_{\eta,\varsigma}}{1-\eta}},1\right\}=1-\beta_{\eta,\varsigma},
\end{align*}
completing the proof. 
\end{proof}

\bibliography{biblio.bib}

@inbook{schelling1966arms,
  author = {Schelling, Thomas C.},
  title = {Arms and Influence},
  series = {The Henry L. Stimson Lectures Series},
  publisher = {Yale University Press},
  year = {1966},
  pages = {74},
  chapter = {1}
}

@article{lipnowski2020repeated,
  title={Repeated delegation},
  author={Lipnowski, Elliot and Ramos, Joao},
  journal={Journal of Economic Theory},
  volume={188},
  pages={105040},
  year={2020},
  publisher={Elsevier}
}

@article{dworczak2019simple,
  title={The simple economics of optimal persuasion},
  author={Dworczak, Piotr and Martini, Giorgio},
  journal={Journal of Political Economy},
  volume={127},
  number={5},
  pages={1993--2048},
  year={2019},
  publisher={The University of Chicago Press Chicago, IL}
}

@article{kolotilin2018optimal,
  title={Optimal information disclosure: A linear programming approach},
  author={Kolotilin, Anton},
  journal={Theoretical Economics},
  volume={13},
  number={2},
  pages={607--635},
  year={2018},
  publisher={Wiley Online Library}
}

@article{chen2008selecting,
  title={Selecting cheap-talk equilibria},
  author={Chen, Ying and Kartik, Navin and Sobel, Joel},
  journal={Econometrica},
  volume={76},
  number={1},
  pages={117--136},
  year={2008},
  publisher={Wiley Online Library}
}

@article{kuvalekar2022goodwill,
  title={Goodwill in communication},
  author={Kuvalekar, Aditya and Lipnowski, Elliot and Ramos, Joao},
  journal={Journal of Economic Theory},
  volume={203},
  pages={105467},
  year={2022},
  publisher={Elsevier}
}

@article{watson1993reputation,
  title={A ``reputation'' refinement without equilibrium},
  author={Watson, Joel},
  journal={Econometrica},
  pages={199--205},
  year={1993}
}

@article{battigalli1997reputation,
  title={On ``reputation'' refinements with heterogeneous beliefs},
  author={Battigalli, Pierpaolo and Watson, Joel},
  journal={Econometrica},
  pages={369--374},
  year={1997},
  publisher={JSTOR}
}

@article{kartik2009strategic,
  title={Strategic communication with lying costs},
  author={Kartik, Navin},
  journal={The Review of Economic Studies},
  volume={76},
  number={4},
  pages={1359--1395},
  year={2009},
  publisher={Wiley-Blackwell}
}

@article{corrao2023bounds,
  title={The Bounds of Mediated Communication},
  author={Corrao, Roberto and Dai, Yifan},
  journal={arXiv preprint arXiv:2303.06244},
  year={2023}
}

@article{
FudenbergLevine89,
author = "Fudenberg, Drew and Levine, David",
title="Reputation and Equilibrium Selection in Games with a Patient Player",
journal="Econometrica",
volume="57",
number="4",
pages="759-778",
year="1989"
}

@article{FudenbergLevine92,
author = "Fudenberg, Drew and Levine, David",
title="Maintaining a Reputation when Strategies are Imperfectly Observed",
journal="Review of Economic Studies",
volume="59",
number="3",
pages="561-579",
year="1992"
}

@article{
Gossner11,
author = "Gossner, Olivier",
title="Simple Bounds on the Value of a Reputation",
journal="Econometrica",
volume="79",
number="5",
pages="1627-1641",
year="2011"
}

@article{kleiner2021extreme,
  author = {Kleiner, Andreas and Moldovanu, Benny and Strack, Philipp},
  title = {Extreme Points and Majorization: Economic Applications},
  journal = {Econometrica},
  volume = {89},
  number = {5},
  pages = {1671--1700},
  year = {2021}
}

@article{arieli2023optimal,
  author = {Arieli, Itai and Babichenko, Yakov and Smorodinsky, Rann and Yamashita, Takuro},
  title = {Optimal Persuasion via Bi-Pooling},
  journal = {Theoretical Economics},
  volume = {18},
  number = {1},
  pages = {15--36},
  year = {2023}
}

@article{ElyValimaki03, 
author = "Ely, Jeffrey and Valimaki, Juuso",
title = "Bad Reputation",
journal = "Quarterly Journal of Economics",
volume = "118",
number = "3",
pages = "785-814",
year = "2003"
}

@article{pei2020interdependent,
  author = {Pei, Harry},
  title = {Reputation Effects Under Interdependent Values},
  journal = {Econometrica},
  volume = {88},
  number = {5},
  pages = {1671--1700},
  year = {2020}
}

@article{liu2011information,
  title={Information acquisition and reputation dynamics},
  author={Liu, Qingmin},
  journal={The Review of Economic Studies},
  volume={78},
  number={4},
  pages={1400--1425},
  year={2011},
  publisher={Oxford University Press}
}

@article{liu2014limited,
  author = {Liu, Qingmin and Skrzypacz, Andrzej},
  title = {Limited Records and Reputation Bubbles},
  journal = {Journal of Economic Theory},
  volume = {151},
  pages = {2--29},
  year = {2014}
}

@article{pei2024shortmemories,
  author = {Pei, Harry},
  title = {Reputation Effects under Short Memories},
  journal = {Journal of Political Economy},
  volume = {132},
  number = {10},
  year = {2024}
}

@article{rayo2010persuasion,
  title={Optimal Information Disclosure},
  author={Rayo, Luis and Segal, Ilya},
  journal={Journal of Political Economy},
  volume={118},
  number={5},
  pages={949--987},
  year={2010}
}

@article{KamenicaGentzkow11,
author="Emir Kamenica and Matthew Gentzkow",
title="Bayesian Persuasion",
journal="American Economic Review",
volume="101",
number="6",
pages="2590-2615",
year="2011"
}

@article{mathevet2024reputation,
  author = {Mathevet, Laurent and Pearce, David and Stacchetti, Ennio},
  title = {Reputation and Information Design},
  year = {2024},
  journal = {Working Paper},
}

@article{fudenberg2022reputation,
  author = {Fudenberg, Drew and Gao, Ying and Pei, Harry},
  title = {A Reputation for Honesty},
  journal = {Journal of Economic Theory},
  volume = {204},
  pages = {105508},
  year = {2022},
  month = {September}
}

@article{best2024persuasion,
  author = {Best, James and Quigley, Daniel},
  title = {Persuasion for the Long Run},
  journal = {Journal of Political Economy},
  volume = {132},
  number = {5},
  year = {2024},
  pages = {1305--1337}
}

@article{ely2008reputation,
  title={When is reputation bad?},
  author={Ely, Jeffrey and Fudenberg, Drew and Levine, David K},
  journal={Games and Economic Behavior},
  volume={63},
  number={2},
  pages={498--526},
  year={2008},
  publisher={Elsevier}
}

@article{myerson1982optimal,
  title={Optimal coordination mechanisms in generalized principal--agent problems},
  author={Myerson, Roger B},
  journal={Journal of mathematical economics},
  volume={10},
  number={1},
  pages={67--81},
  year={1982},
  publisher={Elsevier}
}

@article{morris2001political,
  title={Political correctness},
  author={Morris, Stephen},
  journal={Journal of Political Economy},
  volume={109},
  number={2},
  pages={231--265},
  year={2001},
  publisher={The University of Chicago Press}
}

@article{santambrogio2015optimal,
  title={Optimal transport for applied mathematicians},
  author={Santambrogio, Filippo},
  journal={Birk{\"a}user, NY},
  volume={55},
  number={58-63},
  pages={94},
  year={2015},
  publisher={Springer}
}

@article{pei2023privateLying,
  author = {Pei, Harry},
  title = {Repeated Communication with Private Lying Costs},
  journal = {Journal of Economic Theory},
  volume = {210},
  year = {2023},
  month = {June},
  pages = {105668}
}

@article{rochet1987necessary,
  author = {Rochet, Jean-Charles},
  title = {A Necessary and Sufficient Condition for Rationalizability in a Quasi-Linear Context},
  journal = {Journal of Mathematical Economics},
  volume = {16},
  number = {2},
  year = {1987},
  pages = {191--200}
}

@article{rahman2024detecting,
  author = {Rahman, David M.},
  title = {Detecting Profitable Deviations},
  journal = {Journal of Mathematical Economics},
  volume = {111},
  year = {2024},
  month = {April},
  pages = {102946}
}

@article{matsushima2010role,
  author = {Matsushima, Hitoshi and Miyazaki, Koichi and Yagi, Nobuyuki},
  title = {Role of Linking Mechanisms in Multitask Agency with Hidden Information},
  journal = {Journal of Economic Theory},
  volume = {145},
  number = {6},
  year = {2010},
  pages = {2241--2259}
}

@article{jackson2007overcoming,
  author = {Jackson, Matthew O. and Sonnenschein, Hugo F.},
  title = {Overcoming Incentive Constraints by Linking Decisions},
  journal = {Econometrica},
  volume = {75},
  number = {1},
  year = {2007},
  pages = {241--257},
  month = {January},
  publisher = {The Econometric Society}
}

@article{ball2024quota,
  author = {Ball, Ian and Kattwinkel, Deniz},
  title = {Quota Mechanisms: Finite-Sample Optimality and Robustness},
  journal = {Working Paper}, 
  year = {2024},
  month = {October}
}

@article{escobar2013efficiency,
  author = {Escobar, Juan F. and Toikka, Juuso},
  title = {Efficiency in Games with Markovian Private Information},
  journal = {Econometrica},
  year = {2013},
  volume = {81},
  number = {5},
  pages = {1737--1767},
  month = {September},
}

@article{frankel2014aligned,
  author = {Frankel, Alex},
  title = {Aligned Delegation},
  journal = {American Economic Review},
  volume = {104},
  number = {1},
  year = {2014},
  pages = {66--83},
  month = {January}
}

@article{margaria2018dynamic,
  author = {Margaria, Chiara and Smolin, Alex},
  title = {Dynamic Communication with Biased Senders},
  journal = {Games and Economic Behavior},
  volume = {110},
  year = {2018},
  pages = {330--339},
}

@book{mertens2015repeated,
  title={Repeated games},
  author={Mertens, Jean-Fran{\c{c}}ois and Sorin, Sylvain and Zamir, Shmuel},
  volume={55},
  year={2015},
  publisher={Cambridge University Press}
}

@article{chakraborty2010persuasion,
  author = {Chakraborty, Archishman and Harbaugh, Rick},
  title = {Persuasion by Cheap Talk},
  journal = {American Economic Review},
  volume = {100},
  number = {5},
  year = {2010},
  pages = {2361--2382},
  month = {December}
}

@article{chakraborty2007comparative,
  title={Comparative cheap talk},
  author={Chakraborty, Archishman and Harbaugh, Rick},
  journal={Journal of Economic Theory},
  volume={132},
  number={1},
  pages={70--94},
  year={2007},
  publisher={Elsevier}
}

@article{renault2013dynamic,
  author = {Renault, Jérôme and Solan, Eilon and Vieille, Nicolas},
  title = {Dynamic Sender–Receiver Games},
  journal = {Journal of Economic Theory},
  volume = {148},
  number = {2},
  year = {2013},
  pages = {502--534},
  month = {March}
}

@article{meng2021value,
  author = {Delong Meng},
  title = {On the value of repetition for communication games},
  journal = {Games and Economic Behavior},
  volume = {127},
  pages = {227--246},
  year = {2021},
}

@article{takahashi2010community,
  author = {Satoru Takahashi},
  title = {Community enforcement when players observe partners' past play},
  journal = {Journal of Economic Theory},
  volume = {145},
  number = {1},
  pages = {42--62},
  year = {2010},
}

@article{heller2018observations,
  author = {Yuval Heller and Erik Mohlin},
  title = {Observations on Cooperation},
  journal = {The Review of Economic Studies},
  volume = {85},
  number = {4},
  pages = {2253--2282},
  year = {2018},
  publisher = {Oxford University Press},
}

@article{clark2021record,
  author = {Daniel Clark and Drew Fudenberg and Alexander Wolitzky},
  title = {Record-Keeping and Cooperation in Large Societies},
  journal = {The Review of Economic Studies},
  volume = {88},
  number = {5},
  pages = {2179--2209},
  year = {2021},
}

@article{lin2024credible,
  author = {Xiao Lin and Ce Liu},
  title = {Credible Persuasion},
  journal = {Journal of Political Economy},
  volume = {132},
  number = {7},
  year = {2024},
}

@article{sorin1999merging,
  author = {Sylvain Sorin},
  title = {Merging, Reputation, and Repeated Games with Incomplete Information},
  journal = {Games and Economic Behavior},
  volume = {29},
  pages = {274--308},
  year = {1999},
}

@incollection{avenhaus2002inspection,
  author = {Rudolf Avenhaus and Bernhard von Stengel and Shmuel Zamir},
  title = {Inspection Games},
  booktitle = {Handbook of Game Theory with Economic Applications},
  volume = {3},
  publisher = {Elsevier},
  year = {2002},
  pages = {1947--1987},
  chapter = {51}
}

@article{acemoglu2023mistrust,
  author = {Daron Acemoglu and Alexander Wolitzky},
  title = {Mistrust, Misperception, and Misunderstanding: Imperfect Information and Conflict Dynamics},
  year = {2024},
  journal = {Handbook of the Economics of Conflict}
}

@article{spence1973job,
  author = {Michael Spence},
  title = {Job Market Signaling},
  journal = {The Quarterly Journal of Economics},
  volume = {87},
  number = {3},
  pages = {355--374},
  year = {1973},
  publisher = {Oxford University Press},
}

@book{mailath2006repeated,
  author = {George J. Mailath and Larry Samuelson},
  title = {Repeated Games and Reputations: Long-Run Relationships},
  publisher = {Oxford University Press},
  year = {2006}
}

@article{Arthan2021,
  author = {Rob Arthan and Paulo Oliva},
  title = {On the Borel-Cantelli Lemmas, the Erdős–Rényi Theorem, and the Kochen-Stone Theorem},
  journal = {Journal of Logic and Analysis},
  volume = {13},
  number = {6},
  year = {2021}
}

@book{Rudin1976,
  author    = {Walter Rudin},
  title     = {Principles of Mathematical Analysis},
  edition   = {3rd},
  publisher = {McGraw-Hill},
  year      = {1976},
  address   = {New York}
}

@article{fudenberg2023pathwise,
  author = {Fudenberg, D. and Lanzani, G. and Strack, P.},
  title = {Pathwise Concentration Bounds for Bayesian Beliefs},
  journal = {Theoretical Economics},
  year = {2023},
}

@article{lipnowski2020cheap,
  author = {Lipnowski, Elliot and Ravid, Doron},
  title = {Cheap Talk with Transparent Motives},
  journal = {Econometrica},
  volume = {88},
  number = {4},
  year = {2020},
  month = {July},
  pages = {1631--1660}
}

@article{AlonsoMatouschek2008,
  author = {Ricardo Alonso and Niko Matouschek},
  title = {Optimal Delegation},
  journal = {The Review of Economic Studies},
  volume = {75},
  number = {1},
  year = {2008},
  pages = {259--293},
  publisher = {Oxford University Press},
  url = {https://www.jstor.org/stable/4626195}
}

@book{munkres2000,
  author = {Munkres, James R.},
  title = {Topology},
  publisher = {Prentice Hall},
  year = {2000},
  edition = {2}
}

\end{document}